\documentclass[a4paper,11pt]{article}
\pdfoutput=1

\usepackage{jheppub} 
\usepackage{physics}
\usepackage{graphics}
\usepackage[T1]{fontenc} 
\usepackage{bbold}
\usepackage{dsfont}
\usepackage{multirow}
\usepackage{tikz}
\usepackage{hyperref}  

\DeclareMathOperator{\0}{\mathbb{0}}
\DeclareMathOperator{\1}{\mathbb{1}}
\newcommand{\inspire}{\sc{in}SPIRE}
\newcommand{\Var}{\operatorname{Var}}

\title{\boldmath Detecting scaling in phase transitions on the truncated Heisenberg algebra}

\author[a]{Dragan Prekrat,}
\author[b]{Kristina Neli Todorovi\'{c}-Vasovi\'{c},}
\author[b]{Dragana Rankovi\'{c}}

\affiliation[a]{Faculty of Physics, University of Belgrade, Serbia}
\affiliation[b]{Faculty of Pharmacy, University of Belgrade, Serbia}

\emailAdd{dprekrat@ipb.ac.rs}
\emailAdd{kisi@pharmacy.bg.ac.rs}
\emailAdd{draganat@pharmacy.bg.ac.rs}

\abstract{We construct and analyze a phase diagram of a self-interacting matrix field coupled to curvature of the non-commutative truncated Heisenberg space. The model reduces to the renormalizable Grosse-Wulkenhaar model in an infinite matrix size limit and exhibits a purely non-commutative non-uniformly ordered phase. Particular attention is given to scaling of model's parameters. We additionally provide the infinite matrix size limit for the disordered to ordered phase transition line.}


\begin{document} 
\maketitle
\flushbottom

\section{Introduction}
\label{sec:intro}

Non-commutativity (NC) of space-time was conjured in early days of quantum field theory in hopes of fighting arising infinities \cite{NC1} but soon the magic of renormalization prevailed and NC was forgotten. Since then it was seen to lurk in different corners of physics at different energies, from condensed matter physics to quantum gravity, either as an effective description of encountered phenomena \cite{NC2,NC3} or as a postulated fundamental property of nature. Realization that string theory hides NC at low energies \cite{string1} --- they even appear to share much closer connection \cite{string2} --- finally rekindled the interest for it after many years. But, as if in revenge for abandoning it decades ago, NC cast a severe curse upon field theories on NC spaces: the mixing of UV and IR divergences of non-planar diagrams that damages their renormalizability \cite{UV/IR1,UV/IR2,UV/IR3}.   

Grosse-Wulkenhaar (GW) model \cite{GW1,GW2,GW3,GW4,GW5} is one of rare NC models immune to UV/IR mixing \cite{rNC1,rNC2,rNC3}. It describes a self-interacting real scalar field on the NC Moyal space confined in the
external harmonic oscillator potential. The oscillator term, which shields its renormalizability, can be reinterpreted \cite{tHA} as a coupling with the curvature of the underlining NC space of the truncated Heisenberg algebra $\mathfrak{h}^\text{tr}$. All attempts at generalizing this construction to renormalizable NC gauge models have so far been unsuccessful. 

A common feature of NC field theories is that simultaneously with UV/IR mixing, emerges the translation breaking striped phase in which field oscillates around different values at different points in space and where periodic non-uniform magnetisation patterns appear \cite{striped1,striped2,striped3}. It is believed that this new order lies at the root of UV/IR mixing \cite{MFT}. A while back, we examined a GW inspired gauge model on $\mathfrak{h}^\text{tr}$ space \cite{gauge1,gauge2} which in addition to trivial vacuum possesses another position-dependent one --- a possible hallmark of striped behaviour. Lengthy analytical treatment showed  that divergent non-local derivative counterterms render this model non-renormalizable. We are, in light of this, interested whether the numerical exploration of phases and critical behaviour could indicate nonrenormalizability in advance and save time with future approaches. To that end, in this and the following papers we will numerically compare the behaviour of the matrix regularization of two-dimensional GW model --- whose renormalizability was originally explored in matrix base --- with and without the curvature term. It would be interesting to see if the way the curvature term is turned off affects the limiting phase diagram. This would correspond to the particular way the oscillator term of GW model needs to be turned off by cutoff parameter, in order to assure the two-dimensional NC $\phi^4$-model's renormalizability \cite{GW1}.

Phase diagrams of matrix models on fuzzy spaces have been extensively studied both analytically \cite{MFT,PP,PPP,A00,A01,A02,A03,analytical2,A04,A05,analytical1} and numerically \cite{M2O,N01,N02,N03,N04,N05,N06,N07,triple}. Notable example is $\phi^4$-model on the fuzzy sphere, where we encounter three phases that meet at a triple point. In disordered phase field eigenvalues oscillate around zero, and in ordered phase around one of the two opposite-signed minima of the effective eigenvalue potential. Due to eigenvalue repulsion there is also the third, non-uniformly ordered phase where eigenvalues populate both of these minima. Since we can, in a way, view different eigenvalues as field at different points of space, this phase corresponds to the above-mentioned stripe phase. In fact, there might exist entire series of non-uniformly ordered phases \cite{triple}.

In this paper we analyze in detail the detection of scaling of parameters of each term in the action; this turns out to be nontrivial due to slow convergence and the triple point drifting. We also present the phase diagram for matrices of size $N=24$ and results for infinite matrix size limit of disordered to ordered phase transition line when the interaction with curvature is turned off. 

\section{The model}

The GW model \cite{GW1}
\begin{equation}
S_\text{GW} = \int
\frac{1}{2}(\partial\phi)^2
+ \frac{\Omega^2}{2}((\theta^{-1}x)\phi)^2
+ \frac{m^2}{2}\phi^2
+ \frac{\lambda}{4!}\phi^4,
\end{equation}
with NC embedded in the Moyal-Weyl star product
\begin{equation}
    (f \star g)(x)= \left.\exp (\frac{i\theta^{\mu\nu}}{2}\frac{\partial}{\partial y^\mu}\frac{\partial}{\partial z^\nu})f(y)g(z)\right|_x
    \quad\Longrightarrow\quad
    \comm{x^\mu}{x^\nu}_\star=i\theta^{\mu\nu},
    \quad
\end{equation}
is in \cite{tHA} identified with that of a scalar field coupled with a NC curvature 
\begin{equation}
S_R = \int \sqrt{g}
\left(
\frac{1}{2}(\partial\phi)^2
- \frac{\xi}{2}R\phi^2
+ \frac{M^2}{2}\phi^2
+ \frac{\Lambda}{4!}\phi^4
\right).
\label{Sr}
\end{equation}
The underlying $\mathfrak{h}^\text{tr}$ space satisfies
\begin{equation}
\comm{\mu x}{\mu y} = i\epsilon(1 - \mu z),
\qquad
\comm{x}{z} = i\epsilon\acomm{y}{z},
\qquad
\comm{y}{z} = -i\epsilon\acomm{x}{z},
\end{equation}
where $\epsilon$ is the strength and $\mu$ the mass scale of NC. For $\epsilon=1$, $\mu x$ and $\mu y$ can be represented by finitely-truncated matrices of the Heisenberg algebra
\begin{equation}
X =
\frac{1}{\sqrt{2}}\begin{bmatrix} 
& +1 \\
+1 &  & +\sqrt{2}\\
 & +\sqrt{2} &  & +\sqrt{3}\\
 &  & +\sqrt{3} &  & \ddots\\
 &  &  & \ddots &  & 
\end{bmatrix}_{N\times N}\!\!\!\!\!\!\!\!\!\!\!\!\!,
\qquad
Y =
\frac{i}{\sqrt{2}}\begin{bmatrix} 
& -1 \\
+1 &  & -\sqrt{2}\\
 & +\sqrt{2} &  & -\sqrt{3}\\
 &  & +\sqrt{3} &  & \ddots\\
 &  &  & \ddots &  & 
\end{bmatrix}_{N\times N}\!\!\!\!\!\!\!\!\!\!\!\!\!.
\end{equation}
\!The model \eqref{Sr} was analysed in the frame formalism, with geometry defined by the choice of momenta $p_\mu$ as functions of elements of algebra
\begin{equation}
    \epsilon p_1 = i\mu^2y,
    \qquad
    \epsilon p_2 = -i\mu^2x,
    \qquad
    \epsilon p_3 = i\mu\left(\mu z- \frac{1}{2}\right),
\end{equation}
and derivatives realized as commutators $\partial_\mu f = \comm{p_\mu}{f}$ with these momenta.

\begin{figure}[t]
\centering  
\includegraphics[scale=0.37]{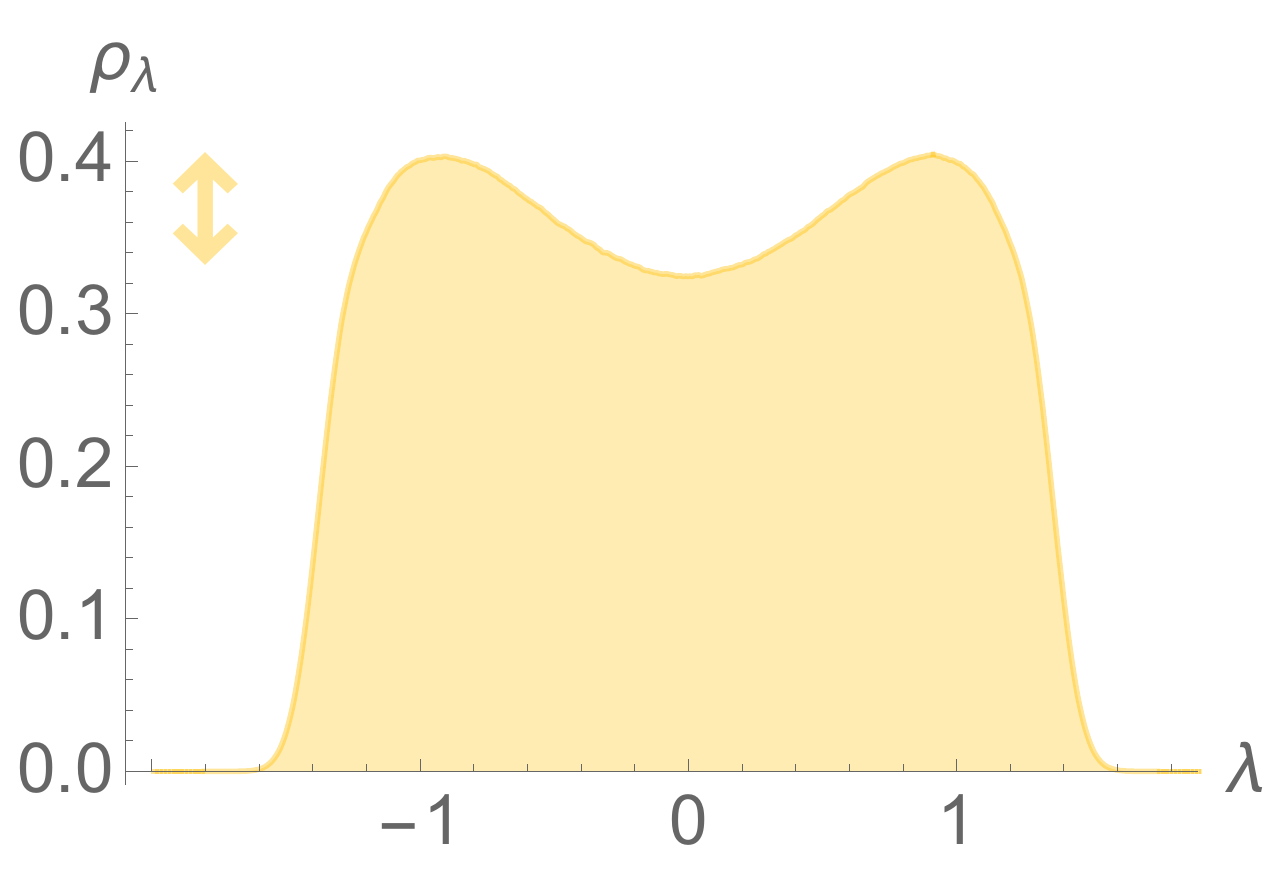}
\includegraphics[scale=0.37]{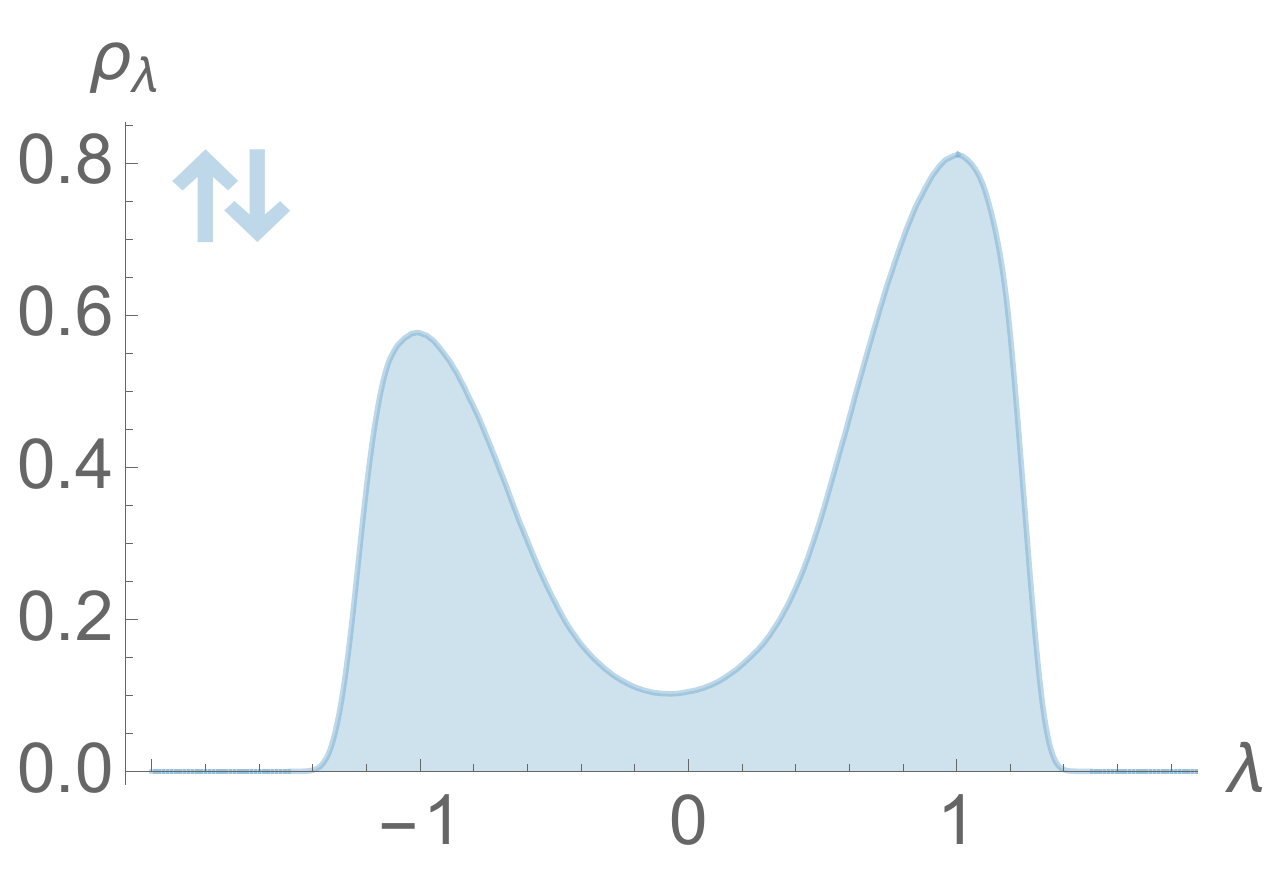}
\includegraphics[scale=0.37]{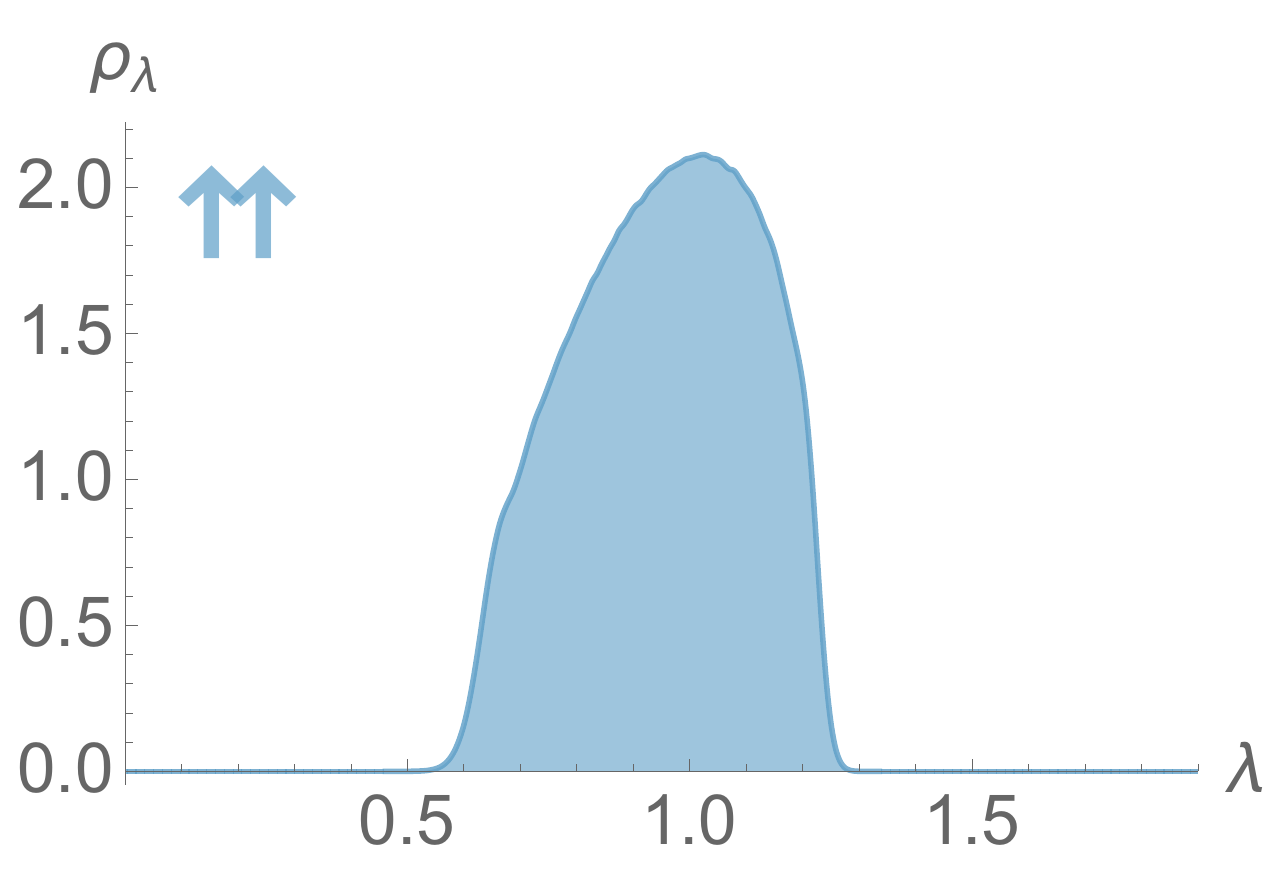}
\vspace{9pt}
\includegraphics[scale=0.37]{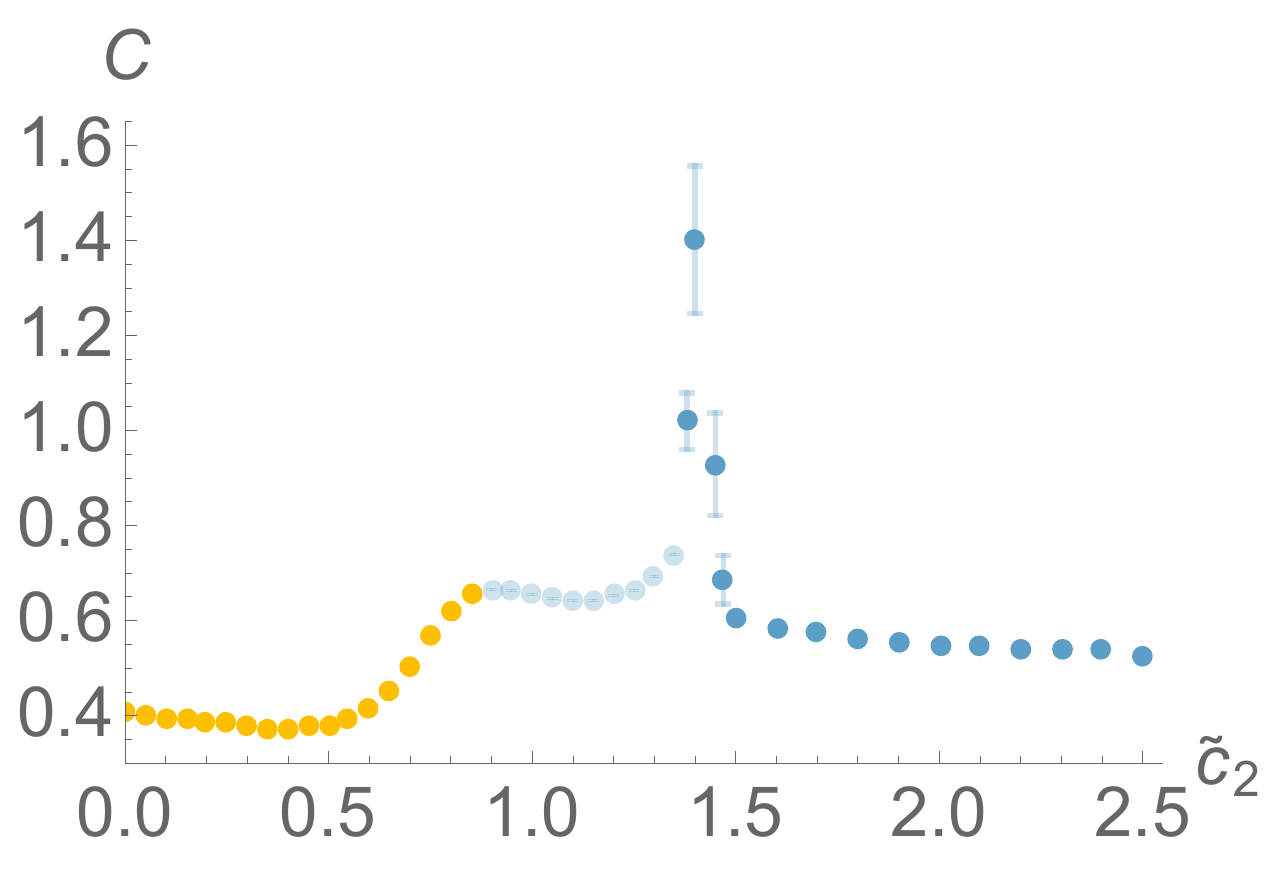}
\includegraphics[scale=0.37]{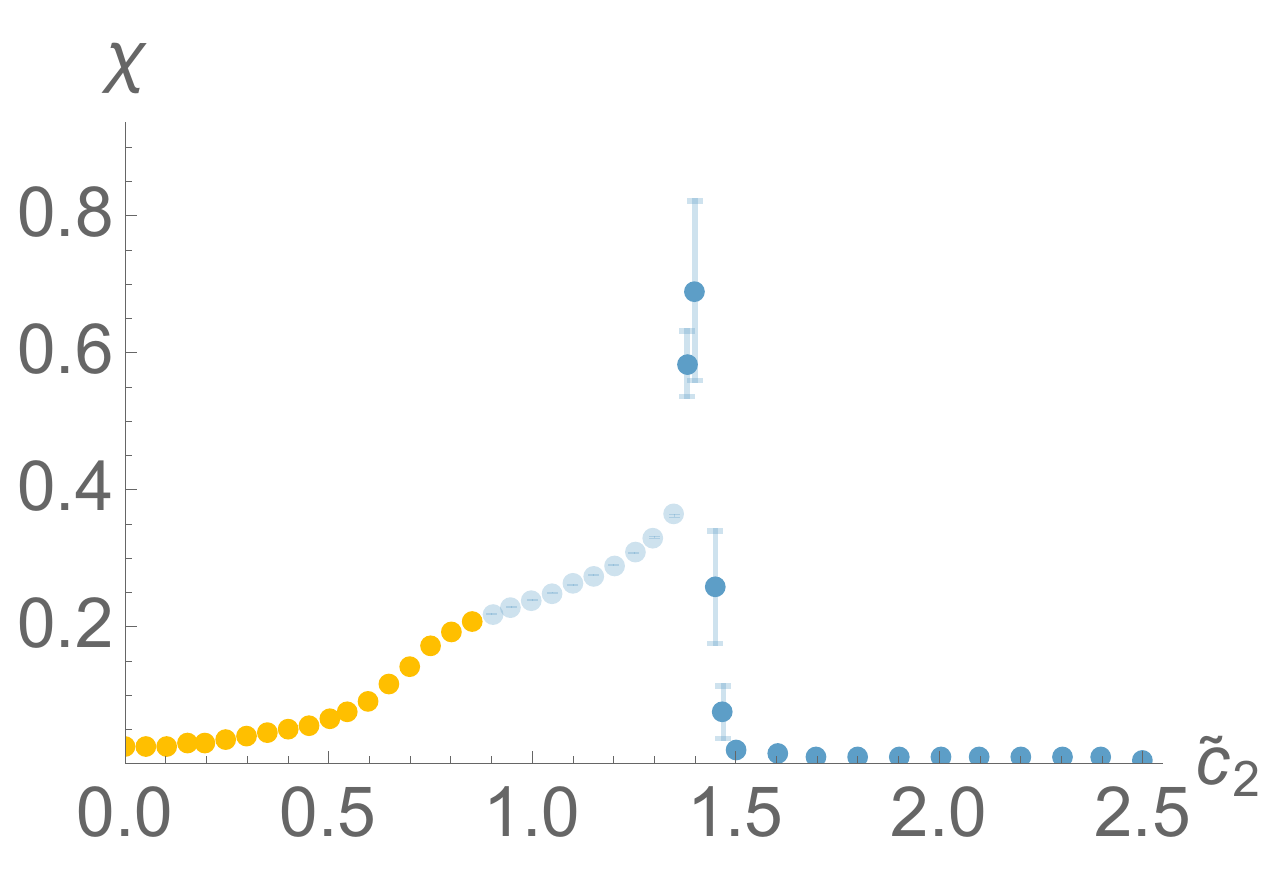}
\includegraphics[scale=0.37]{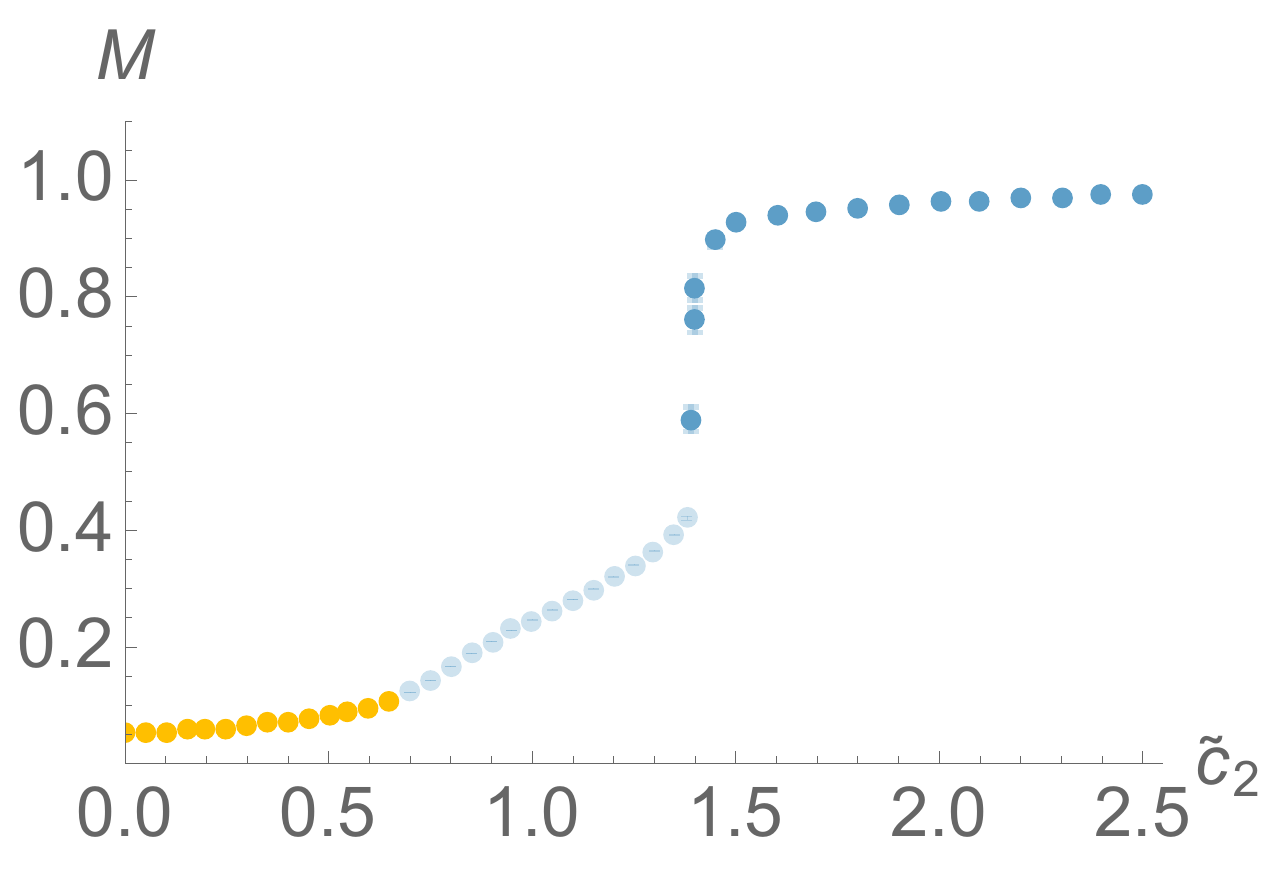}
\vspace{9pt}
\includegraphics[scale=0.37]{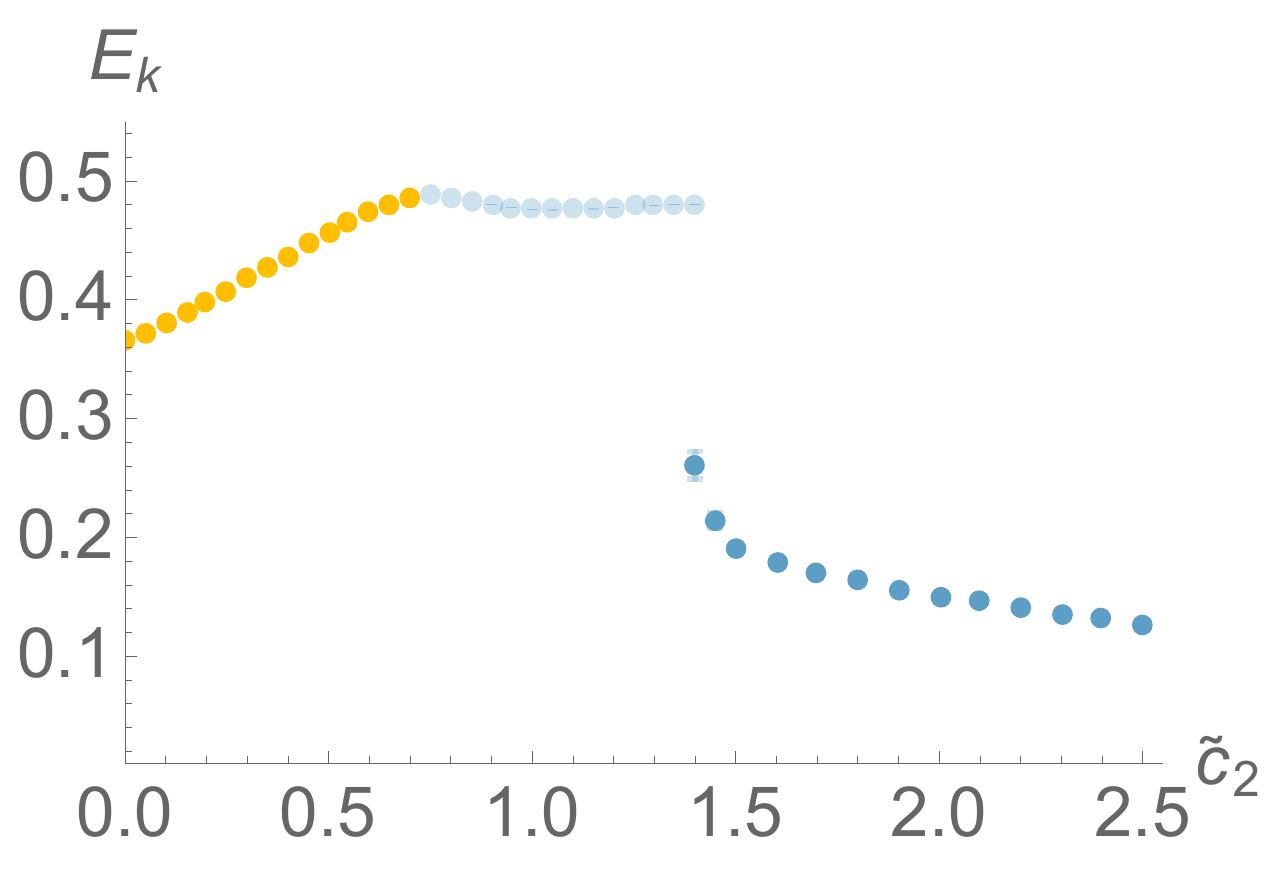}
\includegraphics[scale=0.37]{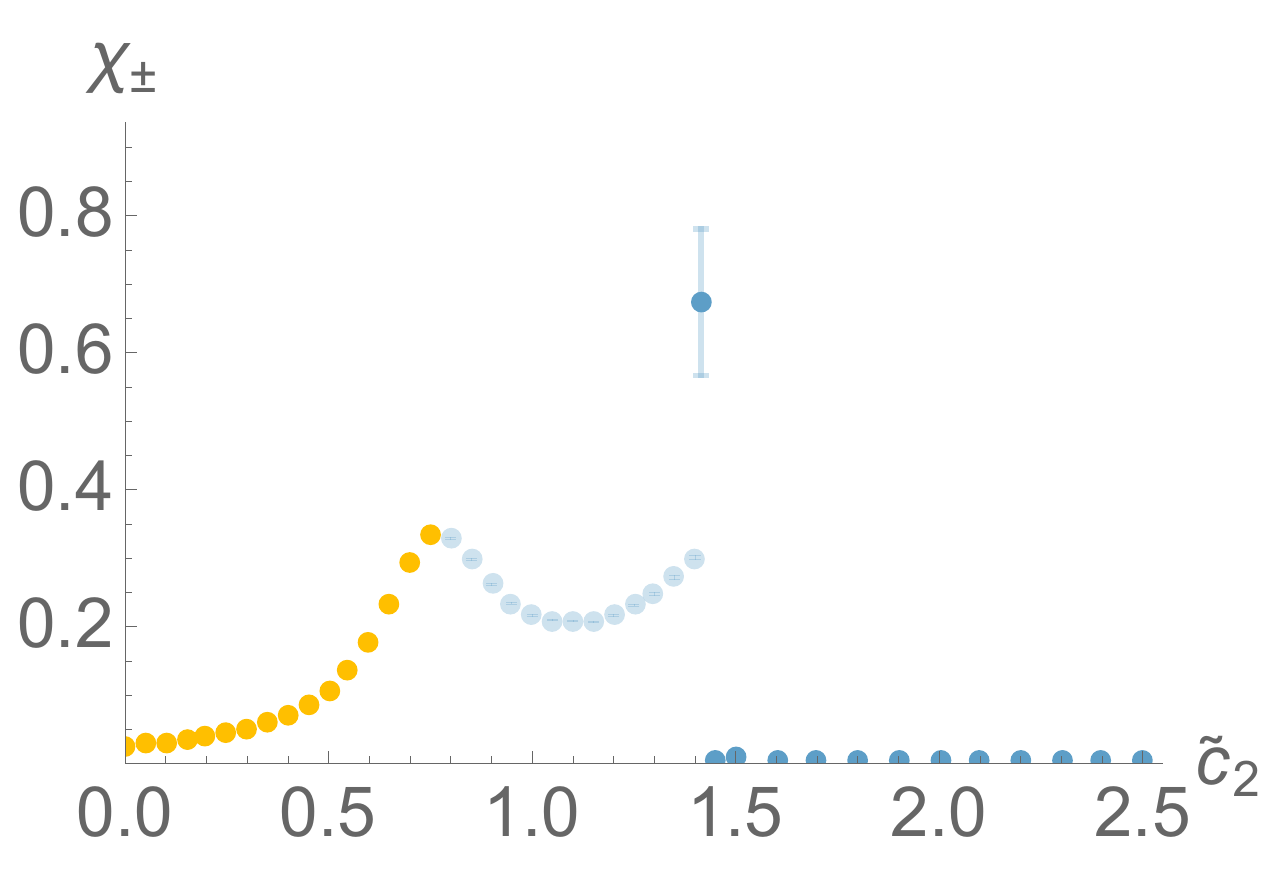}
\includegraphics[scale=0.37]{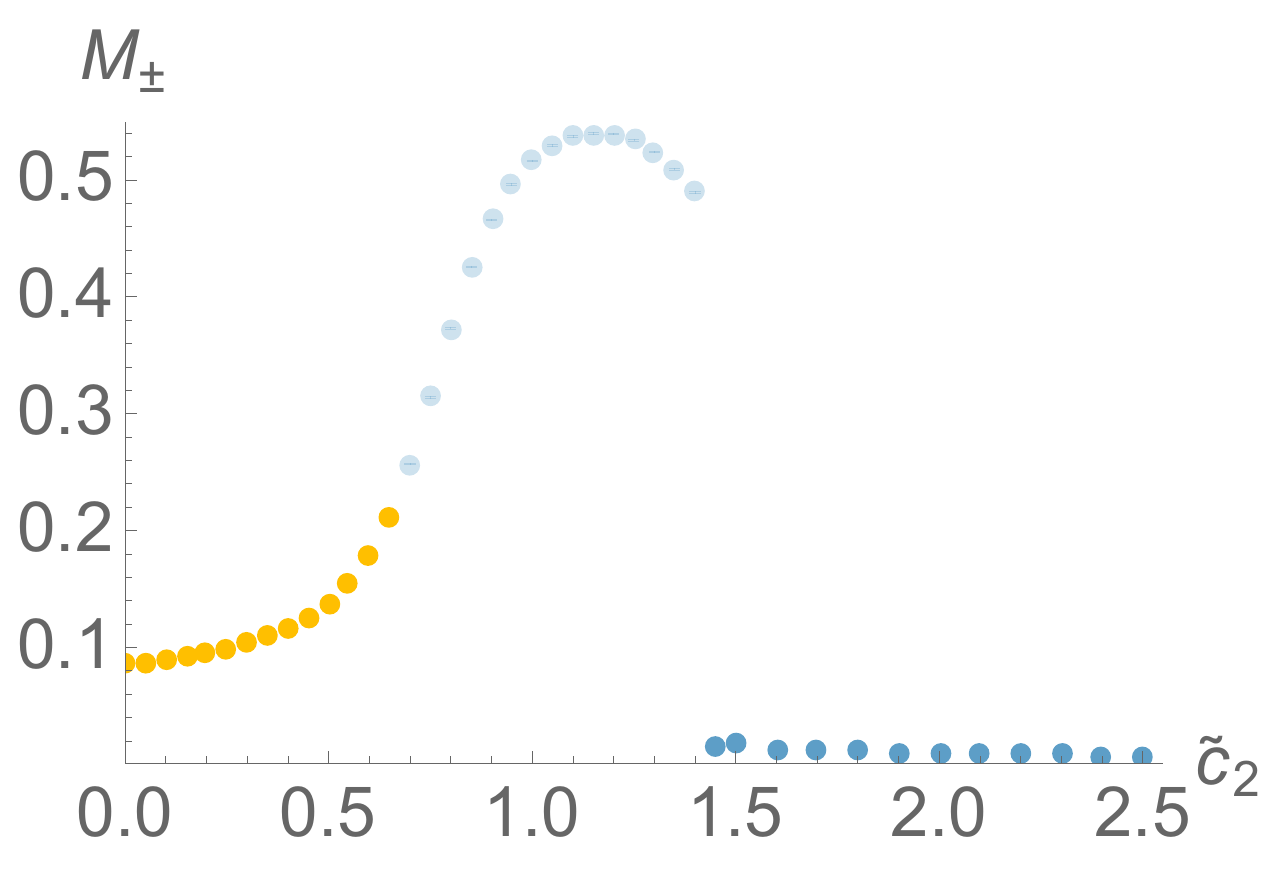}
\vspace{9pt}
\includegraphics[scale=0.37]{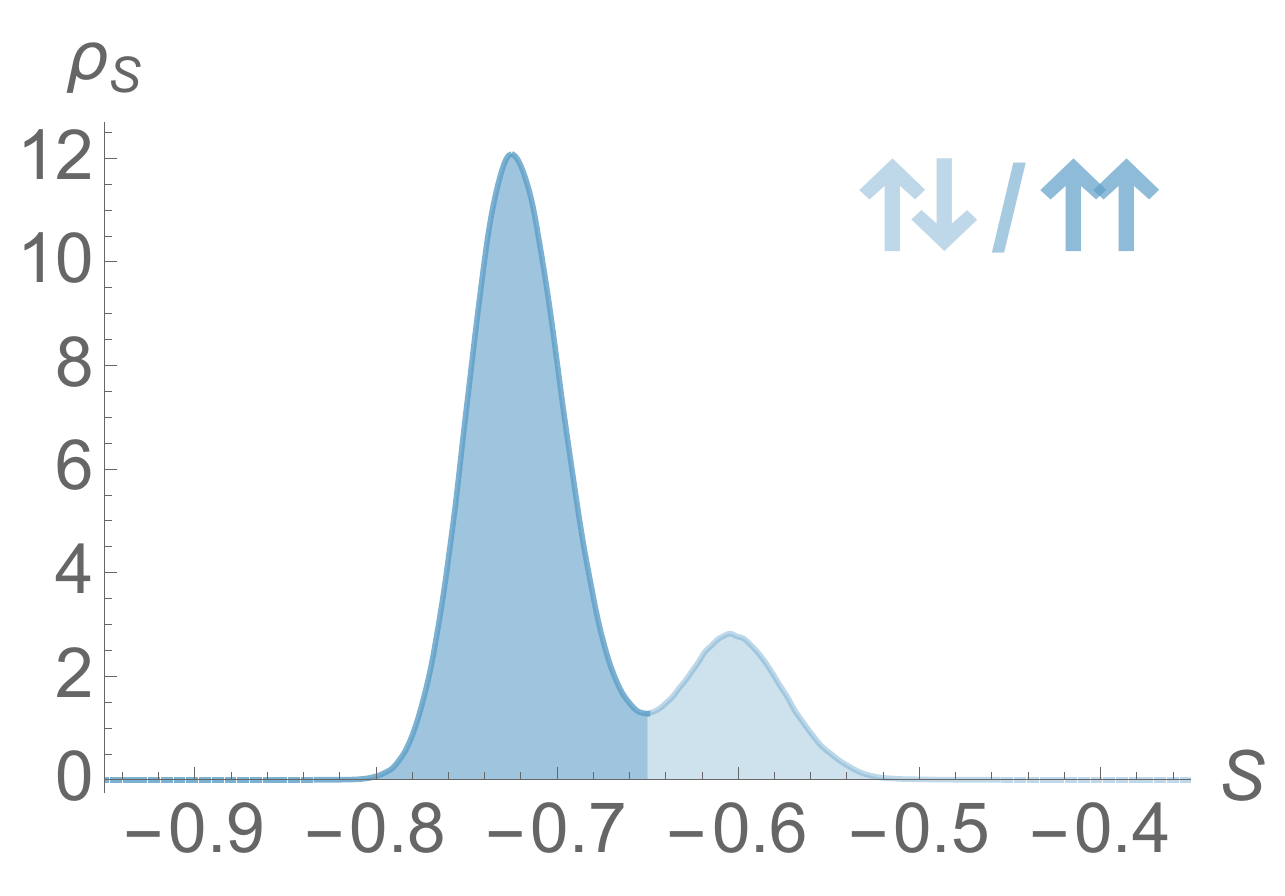}
\includegraphics[scale=0.37]{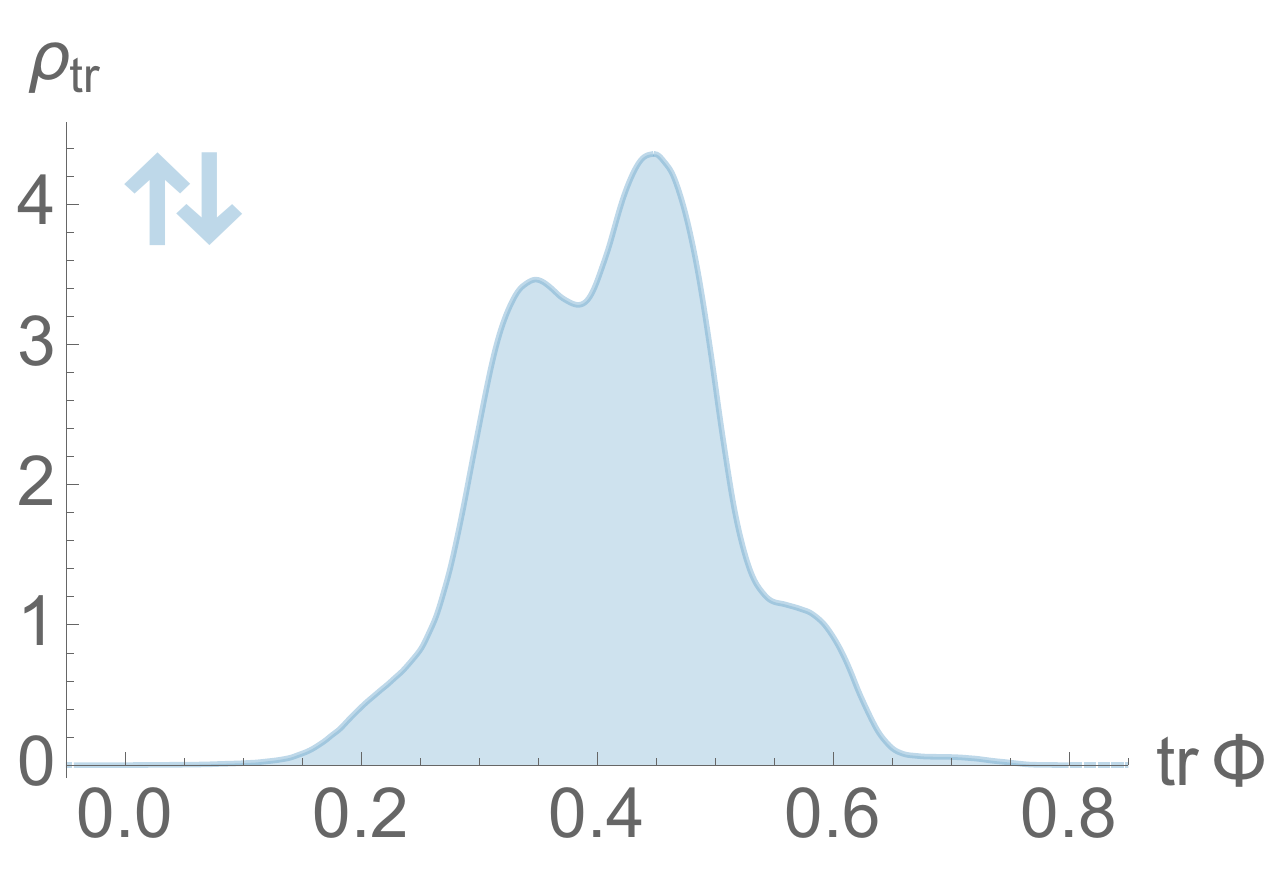}
\includegraphics[scale=0.37]{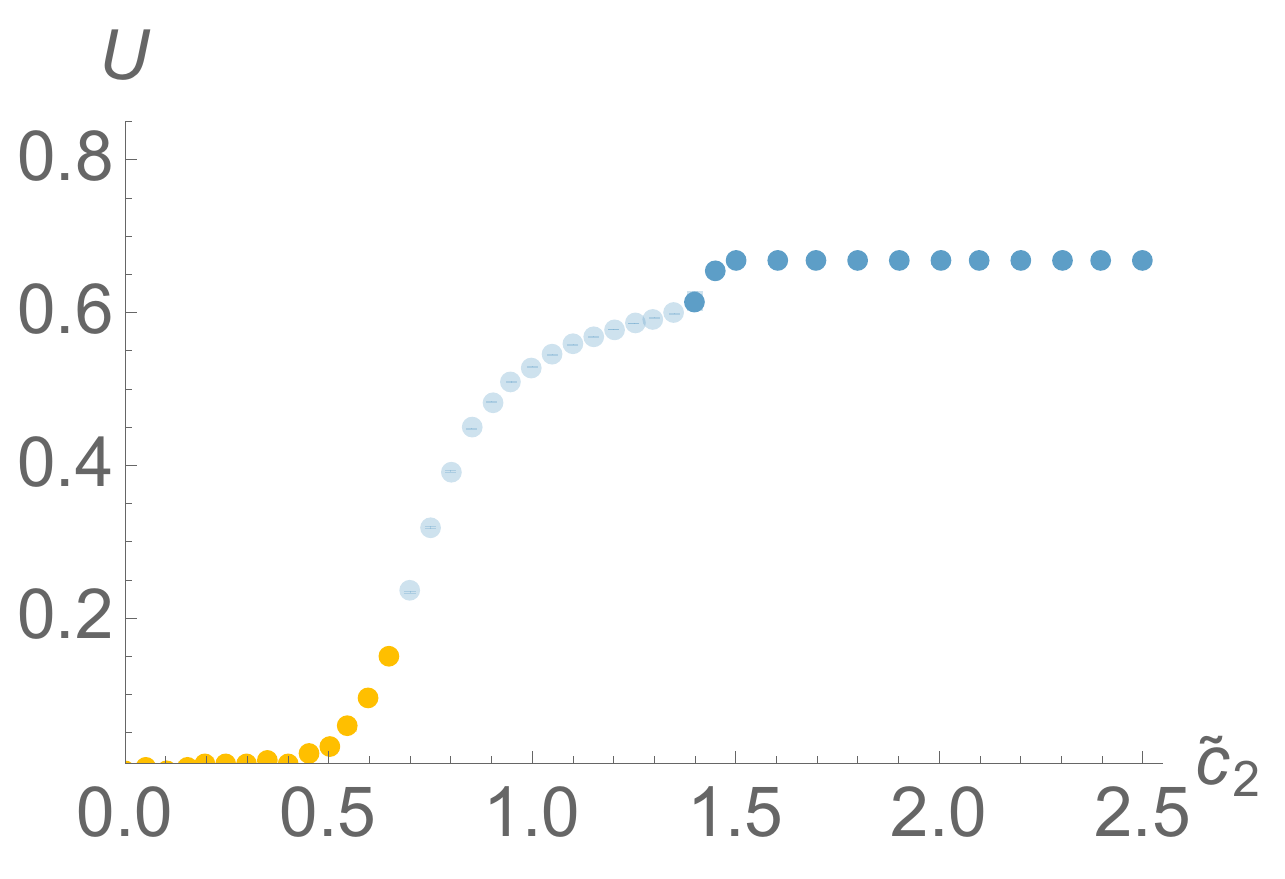}
\caption{Thermodynamic observables for $N=16$, $c_k=1$, $c_r=0$, $\widetilde{c}_4=c_4/N=0.25$, as functions of rescaled mass parameter $\widetilde{c}_2=c_2/N$, with disordered phase colored in yellow/orange and ordered phases in different shades of blue. Transitions are driven by changes in shape of the eigenvalue distribution $\rho_\lambda$, as captured in top row at $\widetilde{c}_2=0.5,\,1.0,\,2.0$ (left to right). We see two transitions as two peaks in $C$ and matching (would-be-) peaks in $\chi$. We also easily see $\updownarrow$ and $\uparrow\uparrow$ phases in plots of $M$ and $U$, while the $\uparrow\downarrow$ phase is clearly visible in staggered magnetization $M_\pm$ and susceptibility $\chi_\pm$ and in $E_k$. Energy distribution $\rho_S$ in the bottom left figure lives at  $\widetilde{c}_2=1.4$, near the border of two ordered phases, and represents two competing states with different energies each belonging to one of the phases. Jump between those states causes 1st order transition and prominent peaks in $C$ and $\chi$. The remaining shy peak in $C$ signals 3rd order transition and it is similar in shape to the well known 3rd order transition of the PP model shown in Figure \ref{Cpp}. Finally, the center bottom figure lives at $\widetilde{c}_2=1.39$ and reveals $\uparrow\downarrow$ phase to be a mixture of different local minimum field configurations with different ratio of positive and negative eigenvalues.  Magnetization and traces are expressed in units of $\sqrt{N\expval{\Tr\Phi^2}}$, eigenvalues in units of $\sqrt{\Tr\Phi^2/N}$, and $S$ in uinits of $\widetilde{c}_2^{\,2}/(4\widetilde{c}_4)$. Errorbars are mostly covered by data markers.}
\label{thermo}
\end{figure}

We investigated a matrix regularization of \eqref{Sr}
\begin{equation}
S_N = \Tr\left(
c_k\Phi\comm{P_\alpha}{\comm{P_\alpha}{\Phi}}
- c_r R\Phi^2
- c_2\Phi^2 + c_4\Phi^4
\right),
\end{equation}
the field $\Phi$ being $N\times N$ hermitian matrix, $P_\alpha$ momenta and $R$ the curvature of $\mathfrak{h}^\text{tr}$ space projected onto $Z=0$ section
\begin{equation}
P_{1}=-Y,
\qquad
P_{2}=X,
\qquad
R=\mathcal{R}\1
-8\left(X^{2}+Y^{2}\right).
\end{equation}
All originally dimensionfull quantities are expressed in units of $\mu$. The minus sign in front of the mass term is chosen for convenience, so that positive $c_2$ parameterizes the relevant portion of the phase diagram.
We used hybrid Monte Carlo, executed in $2^6$ parallel threads each with at least $2^{10}$ decorrelated steps, to measure thermodynamic observables:
\begin{itemize}
    \item energy per degree of freedom $E=\expval{S}\!\big/N^2$,
    \item heat capacity per degree of freedom $C=\Var S\big/N^2$,
    \item magnetization per eigenvalue $M=\expval{|\Tr\,\Phi|}\!\big/N$,
    \item magnetic susceptibility per eigenvalue $\chi=\Var\abs{\Tr\,\Phi}\,\big/N$,
    \item Binder cumulant $U=1-\expval{|\Tr\,\Phi|^4}\!\big/(3\smash{\expval{|\Tr\,\Phi|^2}}^2)$,
\end{itemize}
as well as the control Schwinger-Dyson identity
\begin{equation} \expval{\Tr\left( 2c_k\Phi\comm{P_\alpha}{\comm{P_\alpha}{\Phi}} - 2c_r R\Phi^2
- 2c_2\Phi^2 + 4c_4\Phi^4\right)}=N^2.
\end{equation}
We also kept an eye on the distribution of eigenvalues and traces of the field. Expectation value $\expval{\mathcal{O}}$ and variance $\Var\mathcal{O}$ of the observable $\mathcal{O}$ are given by
\begin{equation}
    \expval{\mathcal{O}} = \frac{\int \dd{\Phi} \mathcal{O} \exp(-S)}{\int \dd{\Phi} \exp(-S)},
    \qquad\qquad
    \Var\mathcal{O}=\expval{\mathcal{O}^2}-\expval{\mathcal{O}}^2.
    \label{observable}
\end{equation}
We computed standard uncertainties $\Delta\mathcal{O}$ from decorrelated data at $68\%$ confidence level. Phase transitions in finite systems manifest as smeared finite peaks and edges in relevant quantities. We scanned through parameter space by varying mass parameter at fixed quatric coupling and searched for peaks in $C$ and $\chi$ (Figure \ref{thermo}). For finite $N$ they do not coincide perfectly, but they converge when matrix size increases. We modeled peaks with triangular distribution of width $w$ and then took $w/(2\sqrt{6})$ as a measure of uncertainty of their position, which gives $65\%$ confidence interval. The edges of the triangular distribution are taken to lie at least $2-3$ standard errors below the best choice for the maximum, with at least $2$ points in proper increasing/decreasing order on the each side of the maximum.

In the absence of kinetic and curvature terms, it is possible to simplify the integration over hermitian matrices in \eqref{observable}, leaving only computationally much cheaper integration over eigenvalues. Since in our case it is not possible to simultaneously diagonalize all four terms, this simplification could not be utilized and we had to settle with working with relatively small matrix sizes. 

Already the analysis of the classical action provides a clue about the structure of the phase diagram. We assume $c_4>0$, to ensure that $S$ is bounded from below. The equation of motion reads
\begin{equation}
2c_k\comm{P_\alpha}{\comm{P_\alpha}{\Phi}}
-c_r\acomm{R}{\Phi}
+ \Phi\left(-2c_2+4c_4\Phi^2\right) = 0,
\end{equation}
and its kinetic, curvature and pure potential parts have respective solutions:\begin{equation}
\Phi = \frac{\Tr\Phi}{N}\1,
\qquad
\Phi = \0,
\qquad
\Phi^2 = 
\begin{cases}
\hspace{5pt} \0 & \textnormal{for } c_2 \leq 0, \\
\displaystyle \frac{c_2\1}{2c_4} & \textnormal{for } c_2 > 0.
\end{cases}
\label{vacuum}
\end{equation}
Obviously, competition is at work between three types of vacua characteristic of three phases discovered in the related matrix models \cite{MFT}:
\begin{itemize}
\item disordered phase: dominant contributions come from oscillations around the trivial vacuum $\expval{\Phi}_{\updownarrow} = \0$,
\item non-uniformly ordered phase (striped phase, matrix phase): dominant contributions come from oscillations around $\expval{\Phi}_{\uparrow\downarrow} \propto \, U\1_\pm U^\dag$, $U$ being a unitary matrix and $\1_\pm$ non-trivial square roots of identity matrix,
\item uniformly ordered phase: dominant contributions from oscillations around $\expval{\Phi}_{\uparrow\uparrow} \propto \1$.
\end{itemize}
We will denote them  $\updownarrow$, $\uparrow\downarrow$ and $\uparrow\uparrow$, respectively.
The pure potential (PP) model, with only mass and quatric term, exhibits the $\updownarrow$ phase for $c_2<0$ and a 3rd order phase transition between $\updownarrow$ and $\uparrow\downarrow$ phases for large enough $c_2>0$. When the kinetic term is turned on, the $\uparrow\uparrow$ phase also appears.

It turns out that the kinetic part of the action $E_k=\expval{S_k}$ and staggered magnetization 
\begin{equation}
M_\pm=\frac{1}{N}\expval{\abs{\Tr((\1_{N/2}\oplus(-\1_{N/2}))\Phi)}}    
\end{equation}
are excellent indicators of the matrix phase: both annihilate highly symmetric $\0$ and $\1$ vacuum states, yielding non-zero contributions on $\1_\pm$. Its accompanying susceptibility is defined as
\begin{equation}
\chi_\pm=\frac{1}{N}\Var\abs{\Tr\,((\1_{N/2}\oplus(-\1_{N/2}))\Phi)}.
\end{equation}

The phases can also be characterised by field's eigenvalue distribution. One-cut deformed Wigner semicircle distribution corresponds to $\updownarrow$ phase and two-cut distribution to $\uparrow\downarrow$ and $\uparrow\uparrow$ phases. However, since eigenvalues come from twin vacuua connected by $\mathbb{Z}_2$-symmetry, for large enough matrices system gets stuck in one of them, so we see asymmetric $\uparrow\downarrow$ and $\uparrow\uparrow$ reduced distributions in Figure \ref{thermo}, accompanied by asymmetric trace distributions. Additionally, Binder cumulant changes sigmoidally with mass parameter, going from $0$ in the $\updownarrow$ phase to $2/3$ in the $\uparrow\uparrow$ phase, deviating into a valley in the $\uparrow\downarrow$ phase (Figure \ref{thermo}).

For the inspected part of parameter space, the $\updownarrow \to \uparrow\downarrow$ transition is visible for $N\ge16$ and the transition to $\uparrow\uparrow$ phase is hard to access (similarly to \cite{M2O}) for values of $c_4$ that allow all 3 phases to occur. The anchoring of the phase diagram is done mostly on the $\updownarrow \to \uparrow\uparrow$ transition line. More details about the transitions, discussion of transition order and critical exponents are provided in the appendix \ref{section:exponents}.

The possibility arises of the novel modification of ordered phases. In the limit of negligible kinetic term, a diagonal solution exists that combines the effects of the curvature and the potential
\begin{equation}
\Phi^2 = \frac{c_2\1+c_r R}{2c_4},
\end{equation}
provided that
\begin{equation}
c_2 \geq \max_j\{c_r \abs{R_{jj}}\}.
\end{equation}
A preliminary analysis of positions of peaks of distribution of eigenvalues and traces seem to corroborate this. We here concentrate mostly on the model without curvature, while the detailed investigation of curvature effects is pending.

\section{Scaling}

Phase diagram of family of models $S_N(c_k,c_r,c_2,c_4;\Phi)$ is expected to converge to a well defined non-trivial large $N$ limit only if we properly choose the scaling of the models' parameters. This allows us to zoom-in on the characteristic features of the diagram as we increase the matrix size.  We will denote scaling of a quantity $q$ with $\nu_q$, so that
$$
q = \widetilde{q}N^{\nu_q},
$$
where 
$\nu_S=2$ stands for the scaling of the action, 
$\nu_\Phi$ for the field/its eigenvalues,
$\nu_P=1/2$ for the momenta,
$\nu_R=1$ for the curvature and 
$\nu_k$,
$\nu_r$,
$\nu_2$,
$\nu_4$ for the coefficients in front of the kinetic, curvature, mass and quatric term respectively.

Requiring each term in the action to behave as $\mathcal{O}(N^2)$ leads, by power counting, to system of equations ($\Tr$ increases power by $1$)
\begin{subequations}
\begin{align}
 \nu_S&=\nu_k+2\nu_P+2\nu_\Phi+1
 \\
 \nu_S&=\nu_r+\nu_R+2\nu_\Phi+1
 \\
 \nu_S&=\nu_2+2\nu_\Phi+1
 \\
 \nu_S&=\nu_4+4\nu_\Phi+1
\end{align}
\end{subequations}
solved by
\begin{align}
 \nu_4=2\nu_2-1,
 \qquad
 \nu_r=\nu_2-\nu_R,
 \qquad
 \nu_k=\nu_2-2\nu_P,
 \qquad
 2\nu_\Phi=1-\nu_2.
\end{align}
For values of $\nu_2$ and $\nu_4$ used in the PP model and on the fuzzy sphere, this amounts to
\begin{equation}
\nu_2=3/2, \quad
\nu_4=2, \quad
\nu_r=1/2, \quad
\nu_k=1/2, \quad
\nu_\Phi=-1/4.
\end{equation}
We wish to examine a simpler choice:
\begin{equation}
\nu_2=1, \quad
\nu_4=1, \quad
\nu_r=0, \quad
\nu_k=0, \quad
\nu_\Phi=0.
\end{equation}
We will also, without loss of generality, set $\widetilde{c}_k=1$, and proceed with the action
\begin{equation}
S_{\text{K}+\text{R}+\text{PP}}(N,\widetilde{c}_2,\widetilde{c}_4,\widetilde{c}_r) = N\Tr\left(\Phi\comm{\widetilde{P}_\alpha}{\comm{\widetilde{P}_\alpha}{\Phi}} 
- \widetilde{c}_r \widetilde{R}\Phi^2 
- \widetilde{c}_2\Phi^2 
+ \widetilde{c}_4\Phi^4\right),
\end{equation}
keeping the rescaled parameters $\widetilde{c}_2$, $\widetilde{c}_4$, $\widetilde{c}_r$ fixed while we increase the matrix size. K stands for the kinetic term, R for the curvature term and PP for the pure potential term.

The wrong choice of scaling would instead of large $N$ stabilization cause the drifting of transition points either towards zero or infinite values in the parameter space. This can be used to identify the correct choice of scaling. It turns out, however, that discriminating between choices based on data is not trivial.

We will first look at the PP term and then see how the kinetic and the curvature terms behave against this well established background.

\section{Pure potential term}

\begin{figure}[t]
\centering
\includegraphics[scale=0.87]{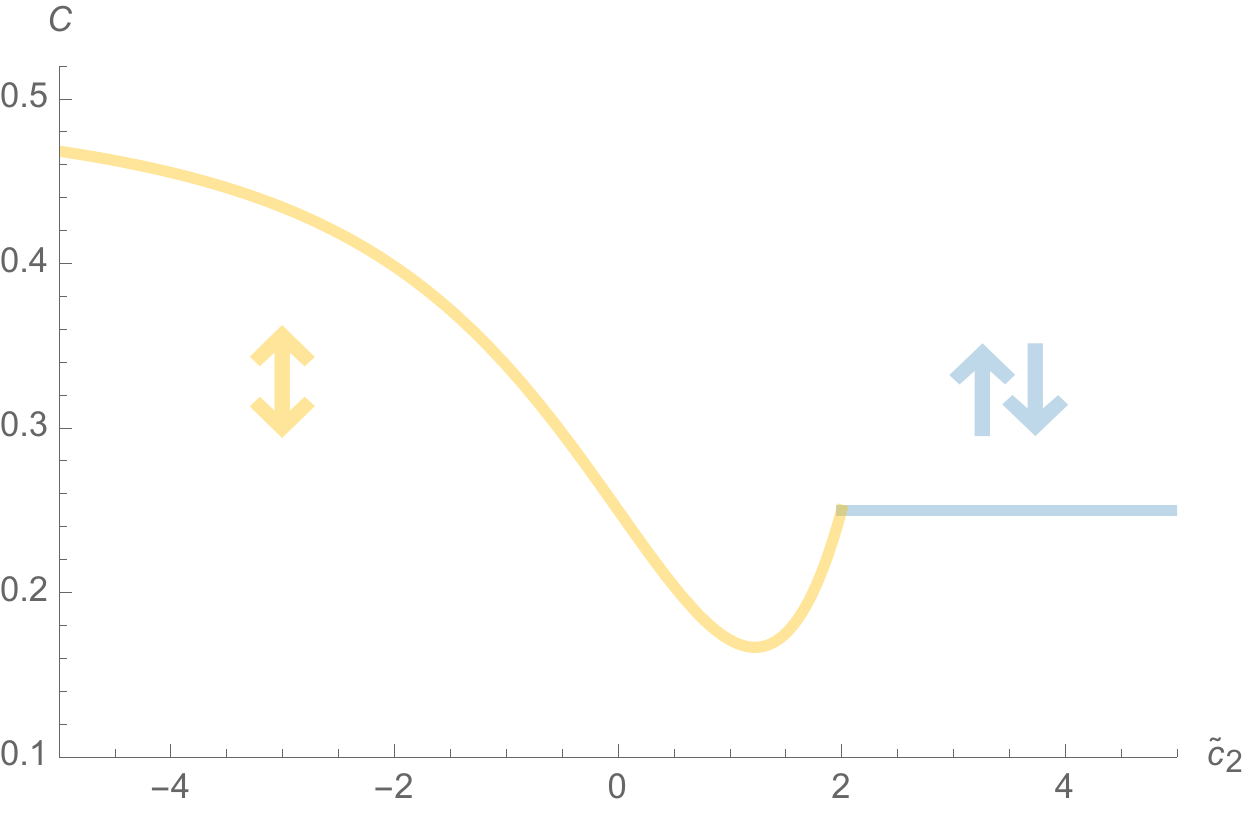}
\caption{3rd order $\updownarrow \to \uparrow\downarrow$ transition in the PP model for $\widetilde{c}_4=1$ in the infinite matrix limit. Derivative of the specific heat has a discontinuity at $\widetilde{c}_2 = 2\sqrt{\widetilde{c}_4}=2$.}
\label{Cpp}
\end{figure}

The PP model 
\begin{equation}
S_\text{PP} = \Tr\left(
- c_2\Phi^2 
+ c_4\Phi^4\right)
\end{equation}
is well studied both analytically and numerically, so it can provide a basic calibration of the method. As it can be seen in Figure \ref{Cpp}, it features a 3rd order transition from $\updownarrow$ to $\uparrow\downarrow$ phase in the large $N$ limit at
\begin{equation}
    c_2=2\sqrt{Nc_4},
\end{equation}
with a sharp-edged kink in specific heat  \cite{A00,PP,PPP}. Both $C$ and $\chi$ remain finite and continuous. At the transition point $C$ reaches value $1/4$ and remains constant for larger $c_2$.

\begin{figure}[t]
\centering
\includegraphics[scale=0.85]{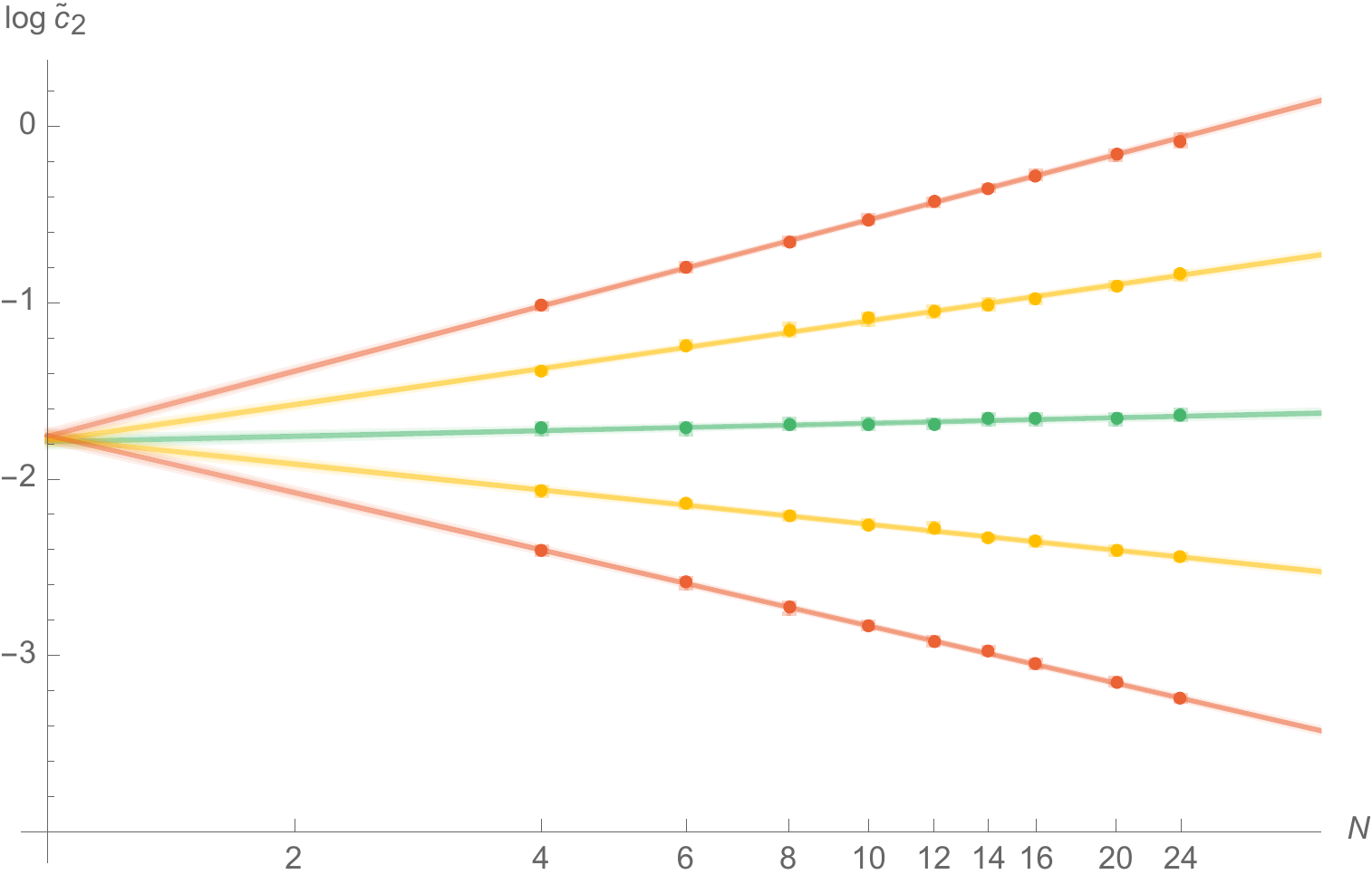}
\caption{$\updownarrow \to \uparrow\downarrow$ transition in the PP model for $\widetilde{c}_4=0.01$, $4\leq N \leq 24$ and fixed $\nu_2=1$, observed as peaks in $\chi$. The green/center data represents the desired choice of scaling $\nu_4=\nu_2=1$, the orange/inner sloped lines $\Delta\nu_4=\pm0.5$ and the red/outer sloped lines $\Delta\nu_4=\pm1$. Pale coloured stripes represent the $68\%$ confidence intervals. Errorbars are mostly covered by data markers.}
\label{PP}
\end{figure}

Transition line equation translates to
\begin{equation}
    \widetilde{c}_2 = 2\sqrt{\widetilde{c}_4N^{1+\nu_4-2\nu_2}}.
    \label{testing}
\end{equation}
Since for desired scaling $\nu_i^*$ phase transition happens at asymptotically fixed rescaled parameters
\begin{equation}
\widetilde{c}_2 = 2\sqrt{\widetilde{c}_4},
\label{c2(c4)PP}
\end{equation}
it must hold
\begin{equation}
    1+\nu_4^*-2\nu_2^*=0.
\end{equation}
Our choice from the previous section satisfies this equality. 
Subtracting this $0$ from the exponent in \eqref{testing}, we get
\begin{equation}
    \widetilde{c}_2 = 2\sqrt{\widetilde{c}_4N^{\Delta\nu_4-2\Delta\nu_2}},
\end{equation}
where $\Delta$ marks the deviation from the desired scaling.
The slope of the logarithmic plot of the transition line equation 
\begin{equation}
\log\widetilde{c}_2 = \frac{\Delta\nu_4-2\Delta\nu_2}{2}\log N + \frac{\log 4\widetilde{c}_4}{2}
\label{log(c2)}
\end{equation}
is therefore changed from zero (up to $\mathcal{O}(1/N)$ effects) to $\Delta\nu_4/2-\Delta\nu_2$, and  Figure \ref{PP} and Table \ref{tab:PP} show how it is affected by different choices of scaling. Both $\nu_2=3/2$, $\nu_4=2$ and $\nu_2=1$, $\nu_4=1$ lead to the correct zero slope and therefore to matrix size independent phase diagram.

That both peaks of $\chi$ and $C$ converge to the same value is demonstrated for $\widetilde{c}_4=0.01$, where the the large $N$ limit of the transition $\widetilde{c}_2$ gives respective values $0.201(8)$ and $0.215(7)$; the theoretical value is $0.2$. 

There is a slight systematic difference (+0.04 on average) between measured and theoretical slopes in Table \ref{tab:PP}. It can be explained as a finite size effect, that disappears for large enough matrices. Namely, since the equation \eqref{c2(c4)PP} is based on the infinite matrix limit, we could account for the finite matrix size by using perturbative ansatz
\begin{equation}
\widetilde{c}_2 = 2\sqrt{\widetilde{c}_4\left(1+\frac{\delta}{\sqrt{N}}+\cdots\right)},
\end{equation}
which modifies \eqref{log(c2)} into
\begin{equation}
\log\widetilde{c}_2 = \frac{\Delta\nu_4-2\Delta\nu_2}{2}\log N + \frac{\log 4\widetilde{c}_4}{2}+\frac{\delta}{2\sqrt{N}}.
\end{equation}
The modified plot is indiscernible from the linear one on the data points, but the intercept and the slope of $\log\widetilde{c}_2-\delta/(2\sqrt{N})$ are perfectly aligned with the theoretical value.  

The results in this section justify the assumption that both conventional and tested choice of scaling are valid, and that there are in fact infinitely many possible ones.

The similar but more nuanced strategy was applied to the curvature term in the appendix \ref{section:curvature} confirming the choice of the chosen parameter scalings.

\begin{table}[t]
\centering
\begin{tabular}{|c|cc|cc|}
    \hline
    \multirow{2}{*}{$\nu_4$} &  \multicolumn{2}{c|}{intercept} & \multicolumn{2}{c|}{slope} \\
    & expected & measured & expected & measured \\
    \hline
    $0.0$ & & $-1.75(4)$ & $-0.50$ & $-0.47(2)$ \\
    $0.5$ & & $-1.77(4)$ & $-0.25$ & $-0.21(2)$ \\
    $1.0$ & $-1.61$ & $-1.79(4)$ & $\phantom{+}0.00$ & $+0.05(2)$ \\
    $1.5$ & & $-1.78(3)$ & $+0.25$ & $+0.30(2)$ \\
    $2.0$ & & $-1.76(4)$ & $+0.50$ & $+0.53(2)$ \\
    \hline
\end{tabular}
\caption{\label{tab:PP} $\log\widetilde{c}_2$ vs. $\log N$ linear fits for $\chi$-transitions for $\widetilde{c}_4=0.01$, $\nu_2=1$ and various $\nu_4$. Differences between expected and measured values are due to finite size effects.}
\end{table}

\section{Kinetic term}
\label{section:K}

Let us now turn on the kinetic term on top of the PP model and consider $S_\text{K+PP}$. As far as transitions go, the action with 
$
(\widetilde{c}_k N^{\Delta\nu_k}, \widetilde{c}_2, \widetilde{c}_4)
$
is equivalent, via absorption of the coefficient into the field, to the one with 
$
(\widetilde{c}_k, \widetilde{c}_2N^{-\Delta\nu_k}, \widetilde{c}_4N^{-2\Delta\nu_k}).
$
Thus would the wrong choice of scaling force the transition points to drift towards zero or the infinity.

\begin{figure}[t]
\centering
\includegraphics[scale=0.90]{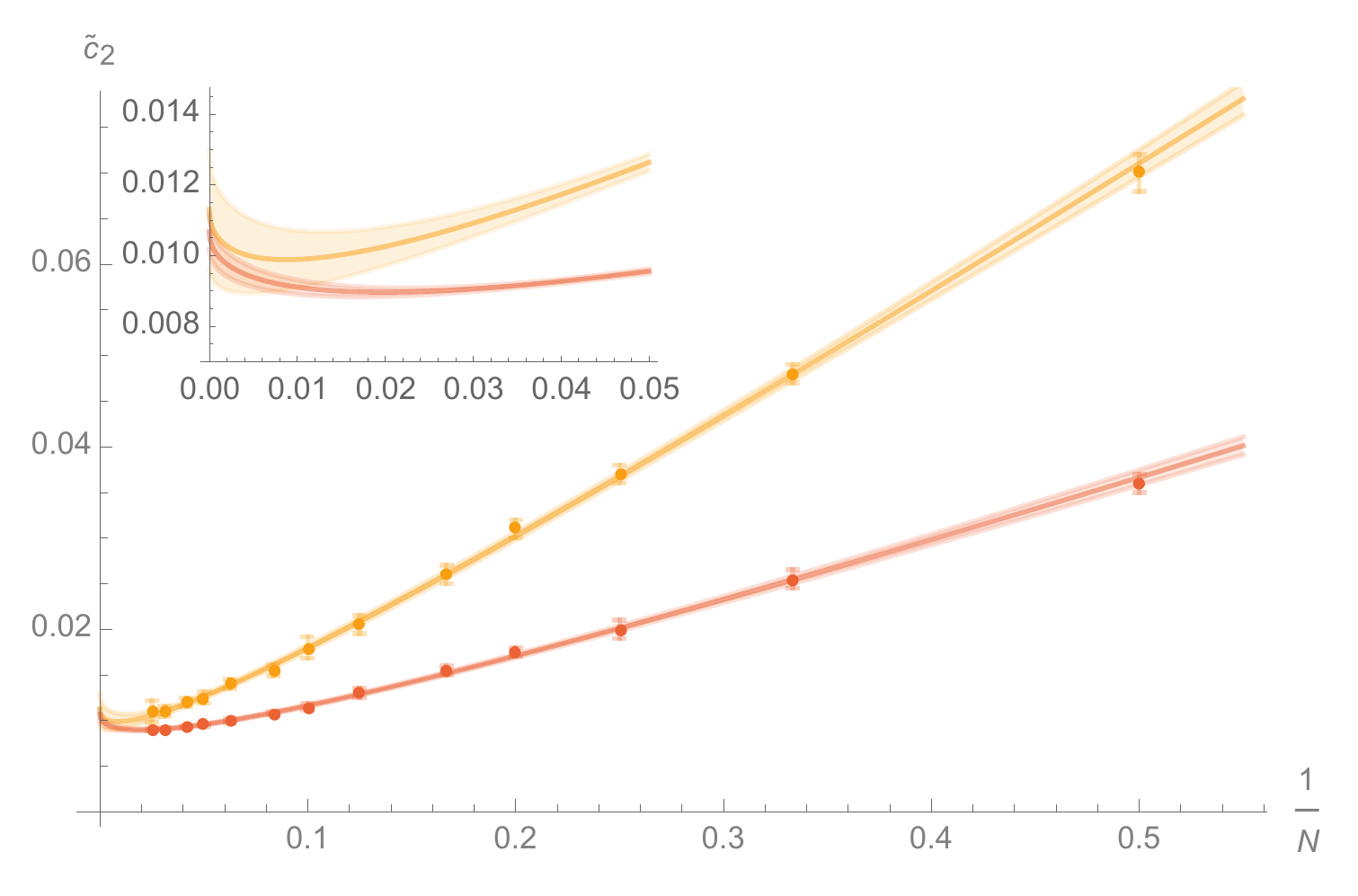}
\caption{$\updownarrow \to \uparrow\uparrow$ transition for $\widetilde{c}_4=0.001$, $\nu_k=0$ and $N\leq40$, observed as peaks in $C$ (orange/top) and $\chi$ (red/bottom). Pale-coloured stripes represent the $68\%$ confidence intervals. The large $N$ limit is zoomed-in.}
\label{convergenceD2O}
\end{figure}

\begin{figure}[t]
\centering
\includegraphics[scale=0.83]{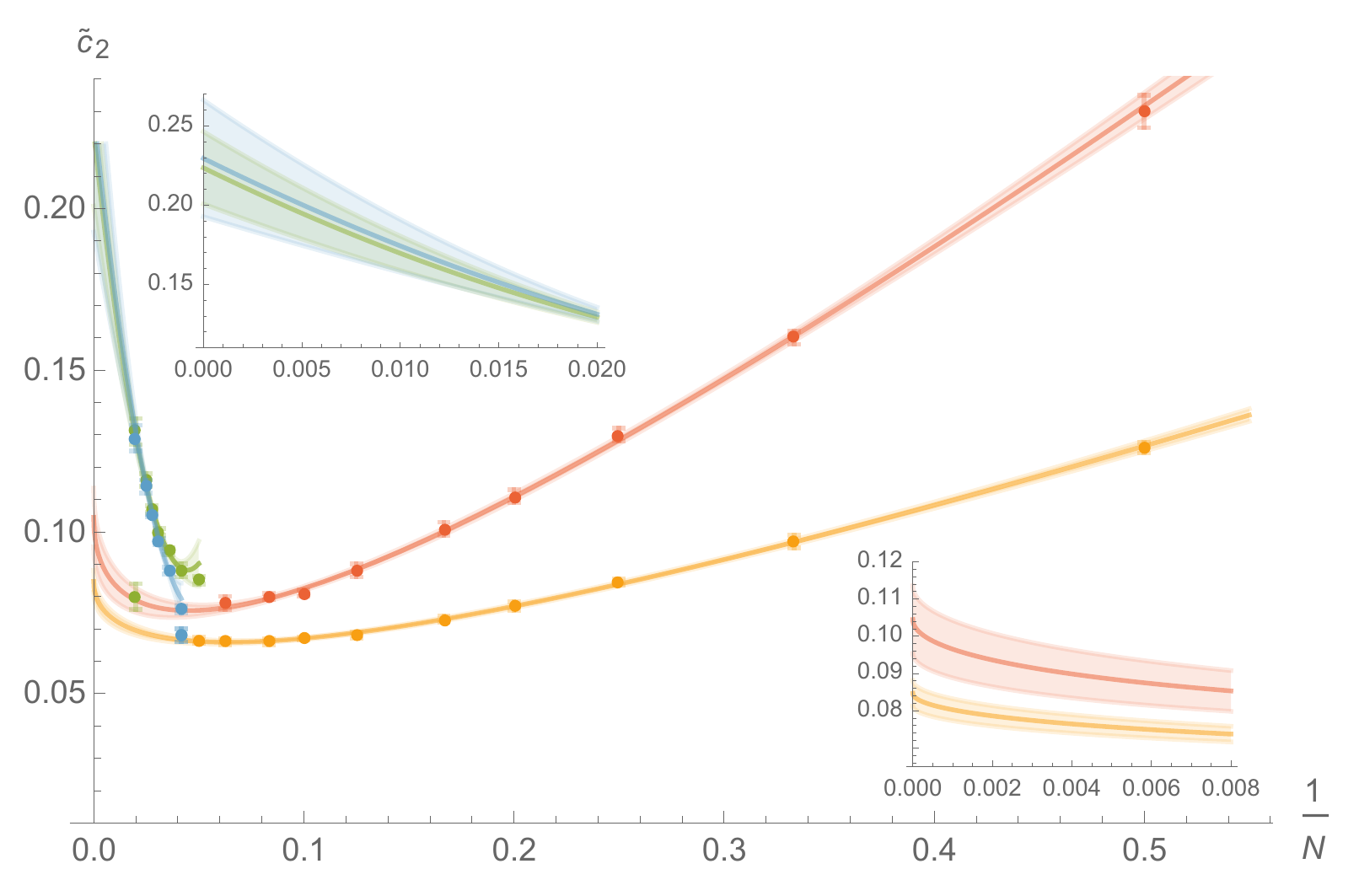}
\includegraphics[scale=0.83]{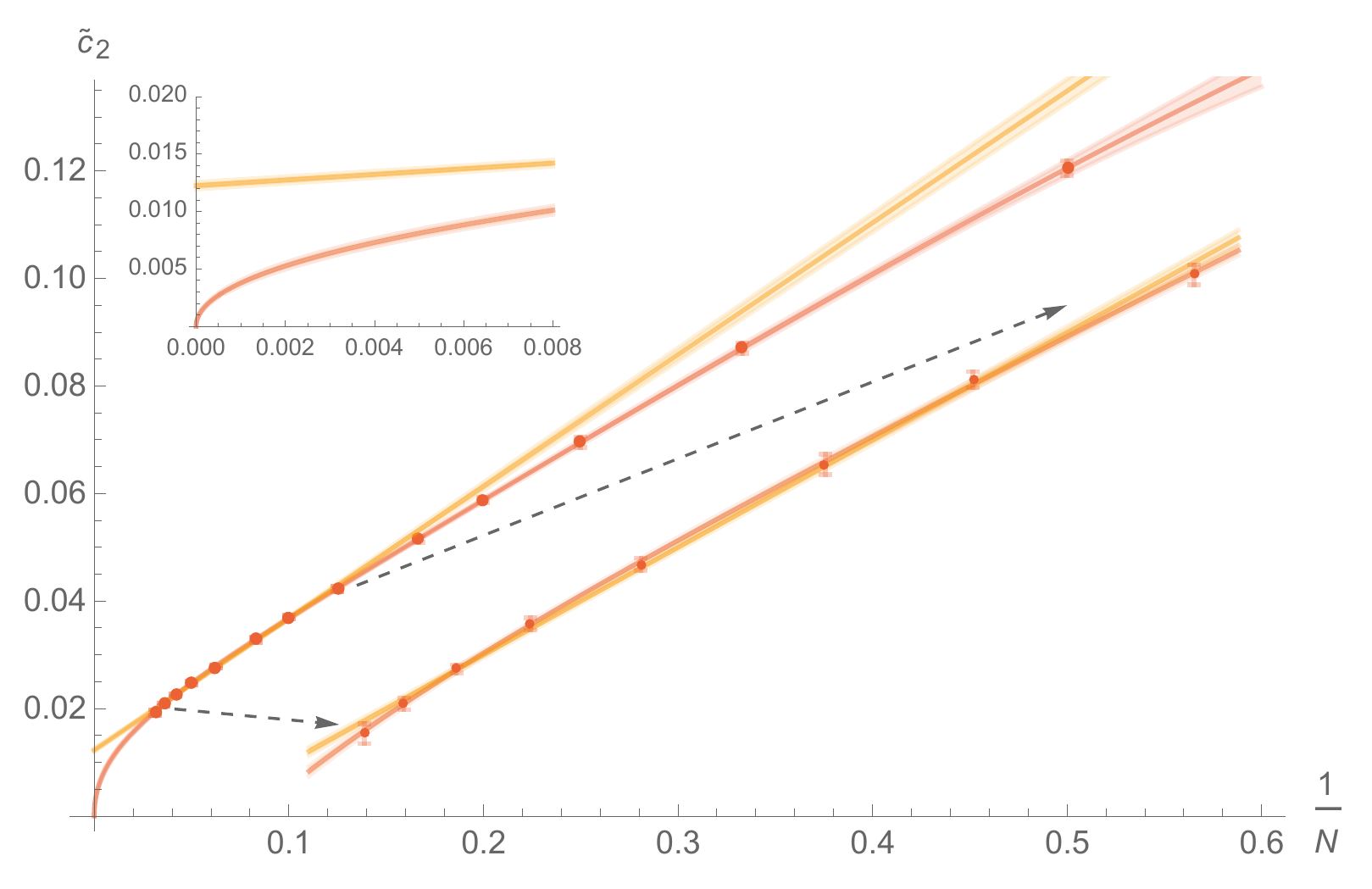}
\caption{{\bf(top)} Transitions for $\widetilde{c}_4=0.01$, $\nu_k=0$, $N \leq 50$ with zoomed-in large $N$ limit. Top plots represents $C$ (red and green) and the bottom ones $\chi$ (orange and blue). $N<16$ is the 2-phase regime (red and orange) and $N>16$ is the 3-phase regime (blue and green). The $\updownarrow \to \uparrow\downarrow$ transition peak fully separates from $\uparrow\downarrow \to \uparrow\uparrow$ peak for $N\geq50$. Pale-coloured stripes represent the $68\%$ confidence intervals. {\bf(bottom)} Transitions for $\widetilde{c}_4=0.01$, $\nu_k=0.5$, $N \leq 32$ with two zoomed-in regions. The orange/top line represents the linear fit for $N \geq 8$, the red/bottom one is our model's prediction. Pale-coloured stripes represent the $68\%$ confidence intervals.}
\label{convergenceM2O}
\end{figure}

The analysis is now complicated by the fact that we lack the analytical prediction for the transition line with kinetic term turned on, so the exact rate of the above mentioned drift is unknown.  Furthermore, discrimination of different scalings based on the data is not clear cut. For example, although Figure \ref{convergenceD2O} shows convincing convergence, looking at the transition plots for $\nu_k=0$ and $\nu_k=0.5$ in Figure \ref{convergenceM2O}, it is not immediately clear which represents the correct choice. At first glance, the wrong choice $\nu_k=0.5$ appears to converge to a non-trivial finite value instead of zero, and the correct choice $\nu_k=0$ to ever increase, possibly towards infinity. One reason for this could be the convergence of the position of the triple point with increasing $N$ closer towards the origin --- the effect demonstrated in \cite{triple} --- causing the system with fixed $\widetilde{c}_4$ to go from 2-phase to 3-phase regime as $N$ increases. The other explanation could be the anomalous negative scaling of the kinetic term, causing the shift towards infinity. Using our data it is not possible to rule out the second option and fix the scaling to precision less than $\pm0.5$, as this would require inspecting much larger matrices. However we can strengthen the case for the choice $\nu_k=0$.

\begin{figure}[t]
\centering
\includegraphics[scale=0.90]{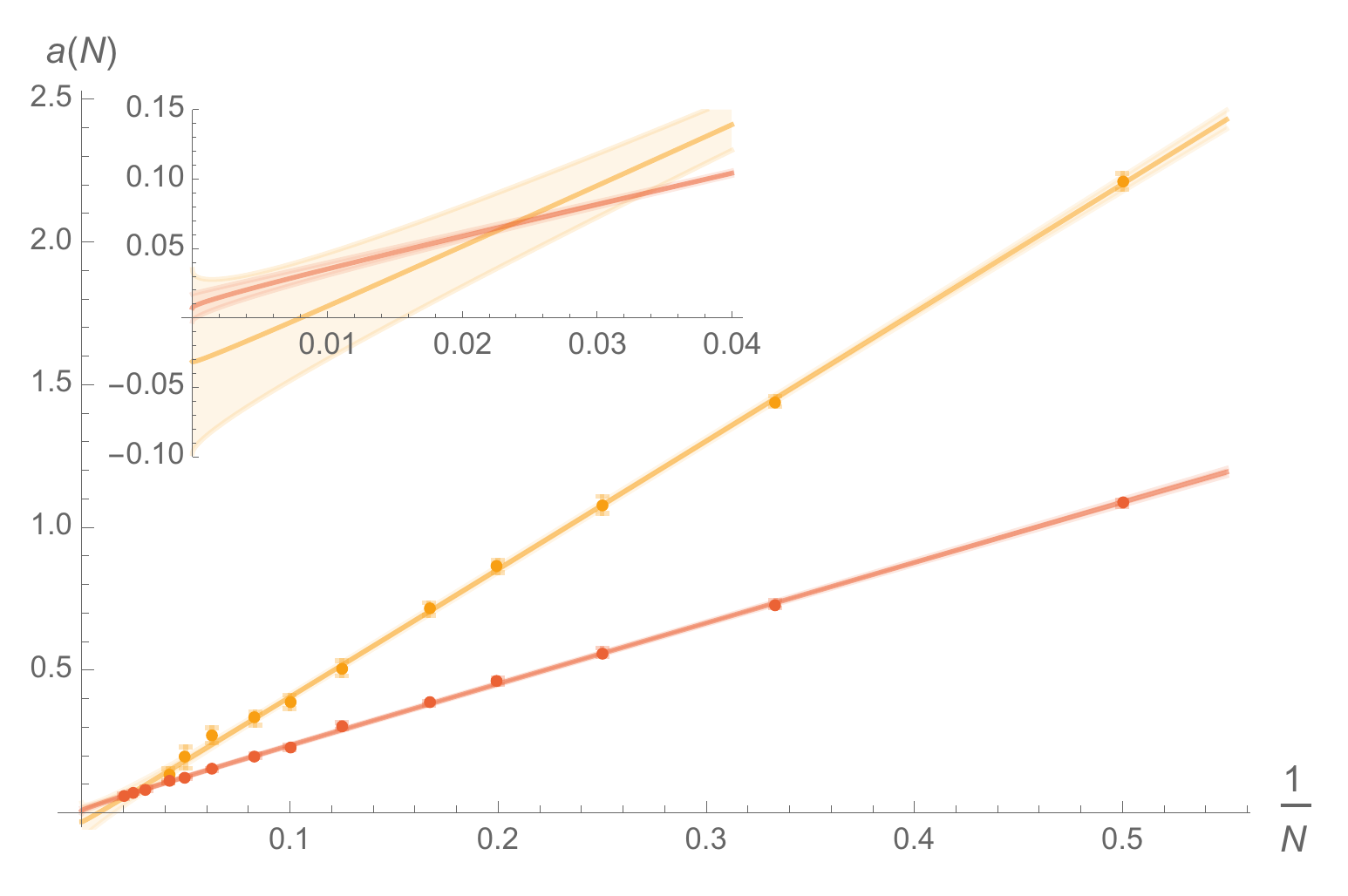}
\includegraphics[scale=0.88]{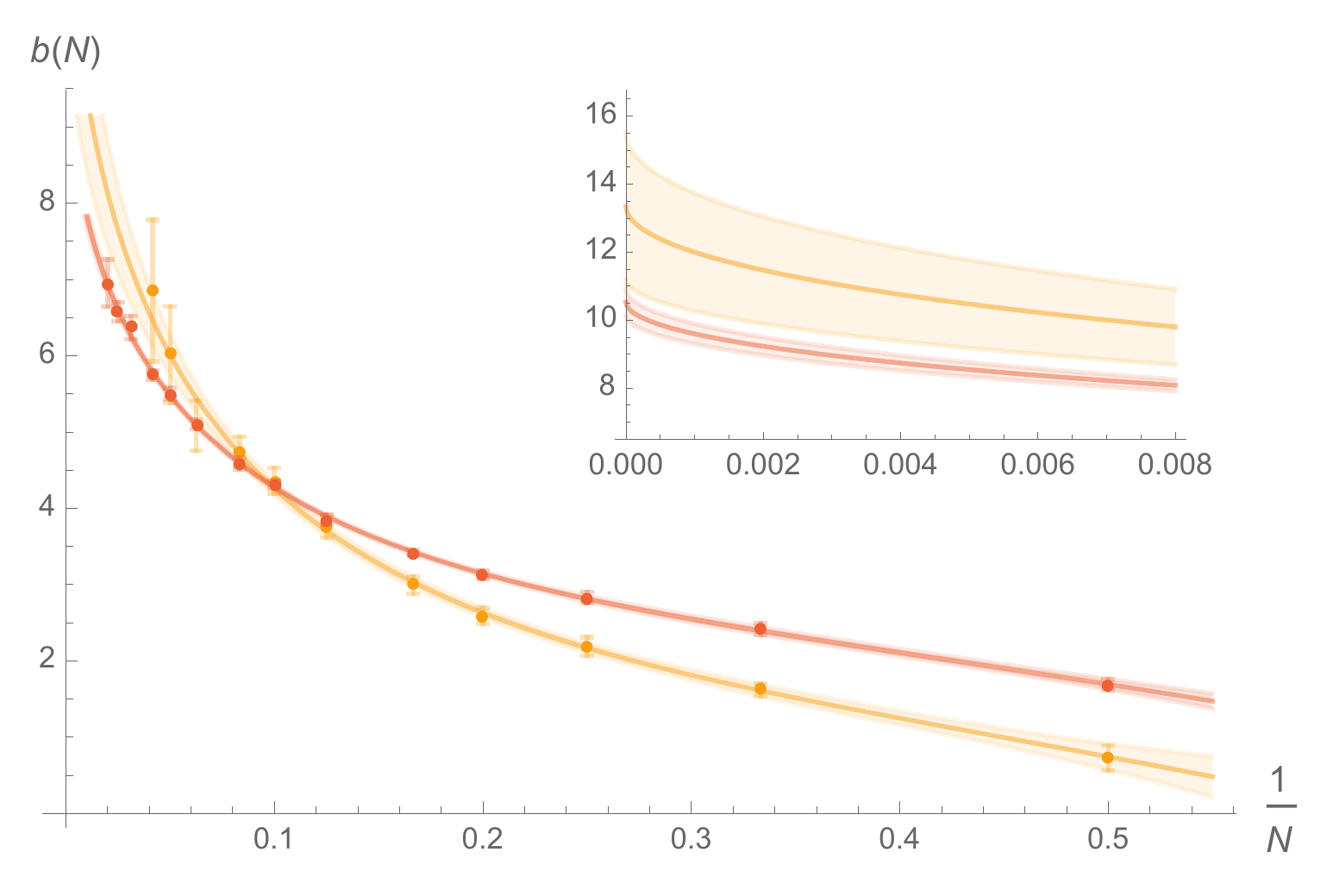}
\caption{$a(N)$ and $b(N)$ coefficients of the $\updownarrow \to \uparrow\uparrow$ transition line constructed from peaks in $C$ (orange/larger errors) and $\chi$ (red/smaller errors) for $N\leq50$. Pale-coloured stripes represent the $68\%$ confidence intervals. The large $N$ limits are zoomed-in. As we can see, the square root behaviour of the transition line, governed by $a(N)$, completely disappears in the infinite matrix limit, leaving only the linear one.}
\label{AB}
\end{figure}

Firstly, Figure \ref{convergenceM2O} (top) allows finite near-linear extrapolation for $1/N\to0$ (in green and blue). Secondly, the change from 2-phase to 3-phase regime for smaller examined $\widetilde{c}_4$ happens at larger $N$, which is consistent with triple point  converging towards smaller $\widetilde{c}_4$. Thirdly, as we will see, extrapolation of the data for $N<16$ (in red and orange) converges to a value consistent with stable linear transition line passing through other smaller values of $\widetilde{c}_4$: had the system not entered 3-phase regime with increasing $N$, the transition line would have passed through $\widetilde{c}_4=0.01$ as well at this extrapolated value of $\widetilde{c}_2$. 

The model on the fuzzy sphere \cite{triple} exhibits linear $\updownarrow \to \uparrow\uparrow$ transition line in the large $N$ limit 
\begin{equation}
\widetilde{c}_2 \propto \widetilde{c}_4.
\end{equation}
In our model, transition for $\nu_k=0$ and fixed $N$ appears to follow the empirical law
\begin{equation}
\widetilde{c}_2 = a(N) \sqrt{\widetilde{c}_4} + b(N)\,\widetilde{c}_4,
\label{c2(c4)}
\end{equation}
where $a(N)$ decreases for larger matrices (Figure \ref{AB}). The coefficients remain stable when higher power of $\widetilde{c}_4$ is added, while the uncertainty makes the higher term indistinguishable from zero.
We are hoping that RG approach \cite{RG1,RG2,RG3} could replicate this form of the transition line; the work on this is currently on the way. 

The wrong choice of scaling would transform \eqref{c2(c4)} into
\begin{equation}
\widetilde{c}_2N^{-\Delta\nu_k} = a(N) \sqrt{\widetilde{c}_4N^{-2\Delta\nu_k}} + b(N)\,\widetilde{c}_4\,N^{-2\Delta\nu_k},
\end{equation}
giving
\begin{equation}
\widetilde{c}_2 = a(N)\sqrt{\widetilde{c}_4} +
b(N)\,\widetilde{c}_4\!
\left(\frac{1}{N}\right)^{\!\!\Delta\nu_k}.
\label{c2(N)}
\end{equation}
We examined several variants of perturbative expansion of $a(N)$ and $b(N)$ as well as a few non-perturbative ones; we did not examine the more complicated possibility that they contain a residual dependence on $\widetilde{c}_4$. The series in $1/\sqrt{N}$ ansatz showed excellent agreement with the collected data: 

\begin{subequations}
\begin{align}
    a(N) & =  \sum_{k=0}^\infty \frac{a_i}{\sqrt{N}^{\,k}} 
    = 0.01(1) + \frac{0.07(7)}{\sqrt{N}} + \frac{2.06(9)}{N},
    \\
    b(N) & = \sum_{k=0}^\infty \frac{b_i}{\sqrt{N}^{\,k}}
    = 10.5(5) - \frac{31(4)}{\sqrt{N}} + \frac{43(9)}{N} - \frac{24(8)}{N\sqrt{N}}.
\end{align}
\label{Puiseux}
\end{subequations}

\noindent The values are confirmed by an analysis of shifts of transition points for different choices of scaling $\Delta\nu_k$ (appendix \ref{section:coefAB}). We also confirmed that the choice of $\nu_k=0$ leads to a stable large $N$ limit. With increasing matrix size $\Delta\nu_k>0$ transition points collapse to zero in the predicted manner which is for $\Delta\nu_k\geq1$ practically linear. 

We can now explain peculiar behaviour of the $\nu_k=0.5$ plot in Figure \ref{convergenceM2O}. Combining \eqref{c2(N)} and \eqref{Puiseux}, we expect
it to change as
\begin{equation}
    \frac{a_1\sqrt{\widetilde{c}_4}+b_0\widetilde{c}_4}{\sqrt{N}} + 
    \frac{a_2\sqrt{\widetilde{c}_4}+b_1\widetilde{c}_4}{N} + 
    \frac{a_3\sqrt{\widetilde{c}_4}+b_2\widetilde{c}_4}{N\sqrt{N}},
\end{equation}
having near constant slope around
\begin{equation}
    N = 3\cdot\frac{a_3\sqrt{\widetilde{c}_4}+b_2\widetilde{c}_4}{a_1\sqrt{\widetilde{c}_4}+b_0\widetilde{c}_4} \approx 3\cdot\frac{b_2}{b_0} = 12(3),
\end{equation}
which falls right in the middle of the observed flat region $8\leq N\leq 32$ on $1/N$ axes, but would ultimately behave as $1/\sqrt{N}$ for large enough matrices.

\section{Phase Diagram}

\begin{figure}[t]
\centering
\includegraphics[scale=0.94]{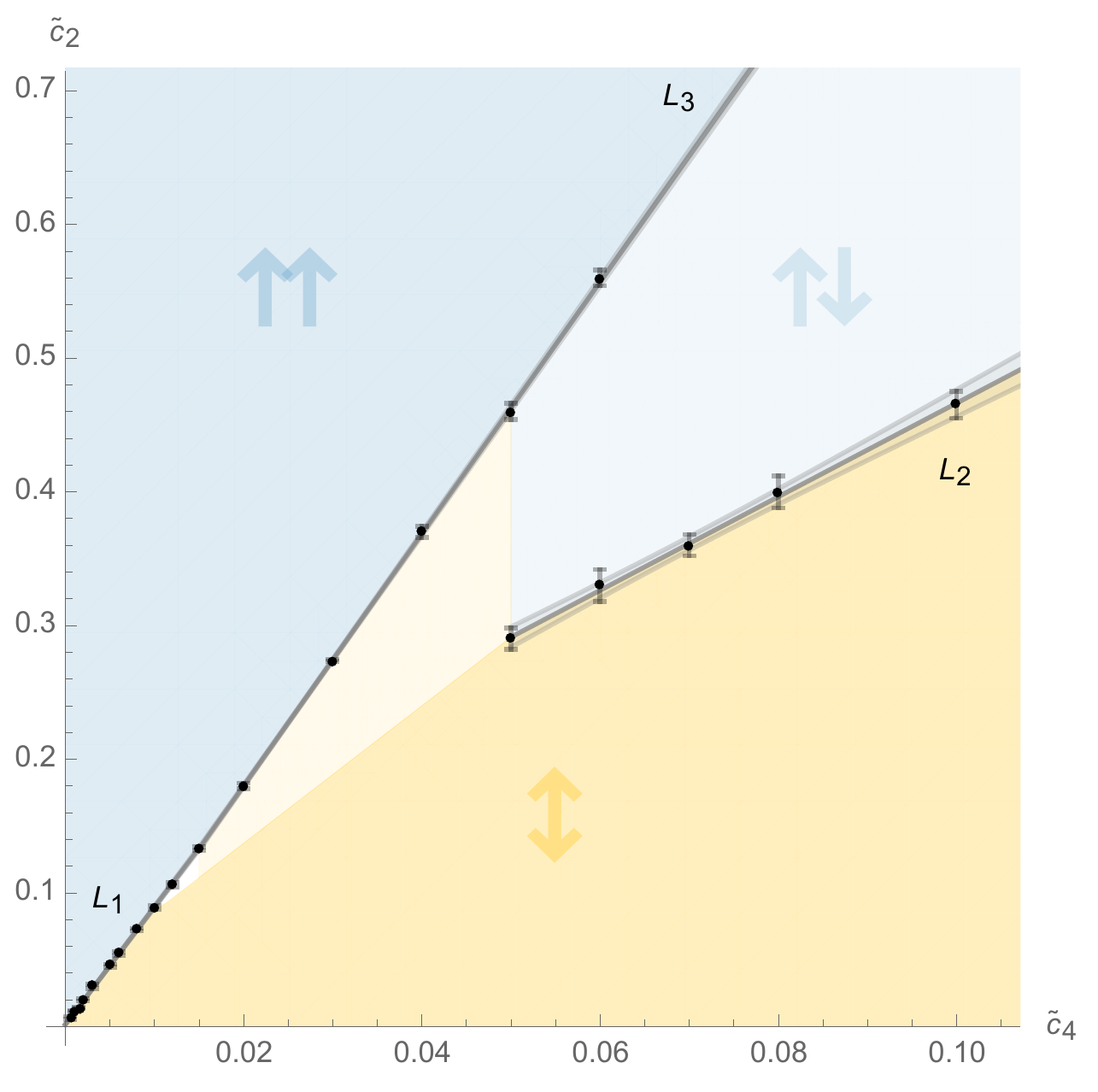}
\includegraphics[scale=0.96]{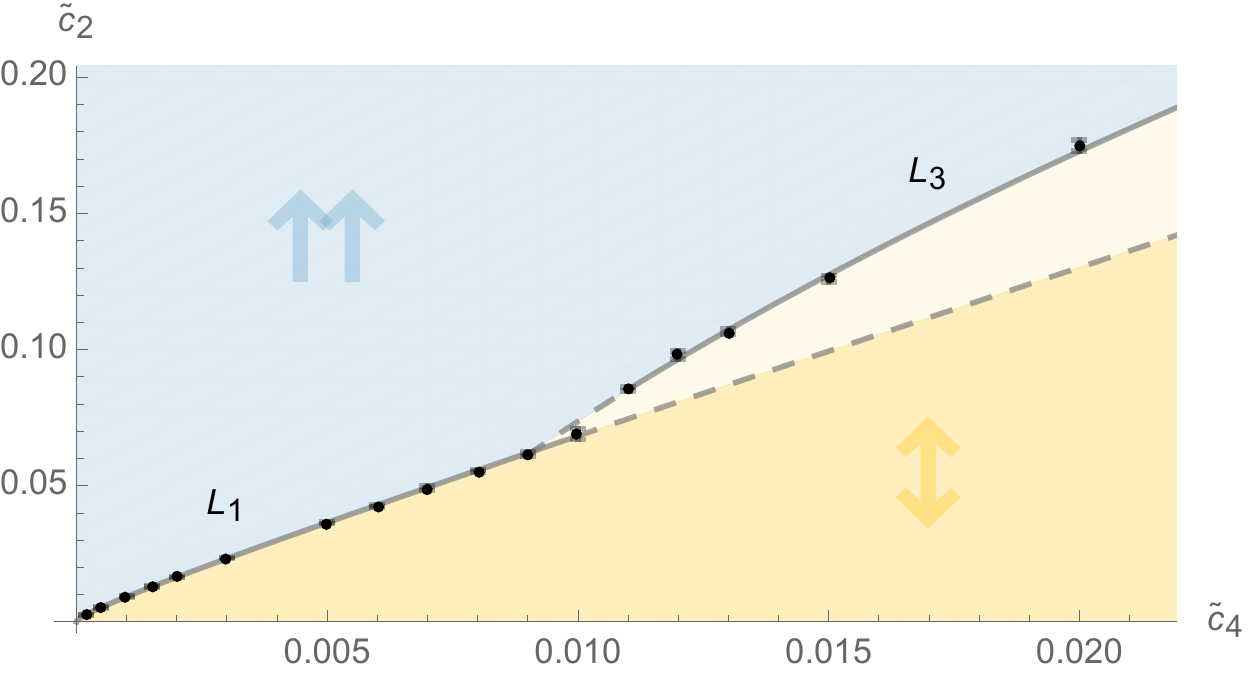}
\caption{Phase diagram for $N=24$. Pale-gray stripes represent $68\%$ confidence intervals. Top diagram is constructed from peaks in $C$ and bottom one from peaks in $\chi$. Bottom plot shows zoomed-in region around the origin of the top plot. A pale yellow wedge between $\updownarrow$ and $\uparrow\uparrow$ phases represents the phase coexistence region, that shrinks with increase in matrix size, and presumably collapses into a triple point.}
\label{N24}
\end{figure}

Having inspected and fixed the scalings, we can finally see how the phase diagram of $S_{\text{K}+\text{PP}}$ model looks like. Figure \ref{N24} depicts the phase structure for $N=24$ obtained from peaks in $C$. From $\widetilde{c}_4=0$ to $\widetilde{c}_4\approx0.015$, stretches the $\updownarrow \to \uparrow\uparrow$ transition line that can be approximated as
\begin{equation}
    L_{1}: \quad \widetilde{c}_2 = 0.0015(4) + 8.8(1)\widetilde{c}_4,
\end{equation}
followed by the $\uparrow\downarrow \to \uparrow\uparrow$ transition line
\begin{equation}
    L_3: \quad \widetilde{c}_2 = -0.009(3) + 9.4(1)\widetilde{c}_4.
\end{equation} 
The slopes of these lines are very similar, making it difficult to determine which points belong to which line; here comes to aid the $\chi$-data in Figure \ref{N24}, clearly showing the transition from $L_1$ to the $L_3$.
Near $\widetilde{c}_4\approx0.05$, $C$-diagram enters a 3-phase regime
and $\updownarrow \to \uparrow\downarrow$ transition line appears, which is linear for smaller $\widetilde{c}_4$
\begin{equation}
    L_2: \quad \widetilde{c}_2 = 0.12(3) + 3.5(3)\widetilde{c}_4,
\end{equation} 
and for larger values of $\widetilde{c}_4$ exhibits square root behaviour characteristic for the limiting PP model
\begin{equation}
    L_2: \quad \widetilde{c}_2 = 2.62(5)\sqrt{\widetilde{c}_4} - 0.48(5) + \frac{0.039(9)}{\sqrt{\widetilde{c}_4}} .
\end{equation}
This can also be seen on the fuzzy sphere \cite{analytical1}, where it holds 
\begin{equation}
    \widetilde{c}_2 = 2.5\sqrt{\widetilde{c}_4}+\frac{0.5}{1-\exp(1/\sqrt{\widetilde{c}_4})} 
    \approx 
    2\sqrt{\widetilde{c}_4} + 0.25 - \frac{0.042}{\sqrt{\widetilde{c}_4}}.
\end{equation}
It would be interesting to compare these two once the large $N$ extrapolation of the $L_2$ is obtained. A very crude linear extrapolation of $N=16,20,24$ gives promising $2.0(4)$ for the square root coefficient. 

The extrapolation of $L_2$ intersects $L_{1/3}$ at $\widetilde{c}_4\approx0.02$,  which is in the vicinity of the meeting point of $L_1$ and $L_3$ at $\widetilde{c}_4\approx0.015$ ($\widetilde{c}_4\approx0.01$ for $\chi$-data), placing the would-be triple point nearby. The pale yellow triangle formed by the meeting point of $L_1$ and $L_3$ and the starting point of $L_2$ should collapse into a triple point when $N\to\infty$. This effect is in fact demonstrated on the fuzzy sphere \cite{triple}. In this region the two transition peaks are still convoluted into a single one (like peaks of $\chi$ in Figure \ref{thermo}). 

Expression for  $L_3$ should be taken with a grain of salt. This is where the ergodicity of algorithm starts to falter, contributing to an unknown systematic error. 
\begin{figure}[t!]
\centering
\includegraphics[scale=0.90]{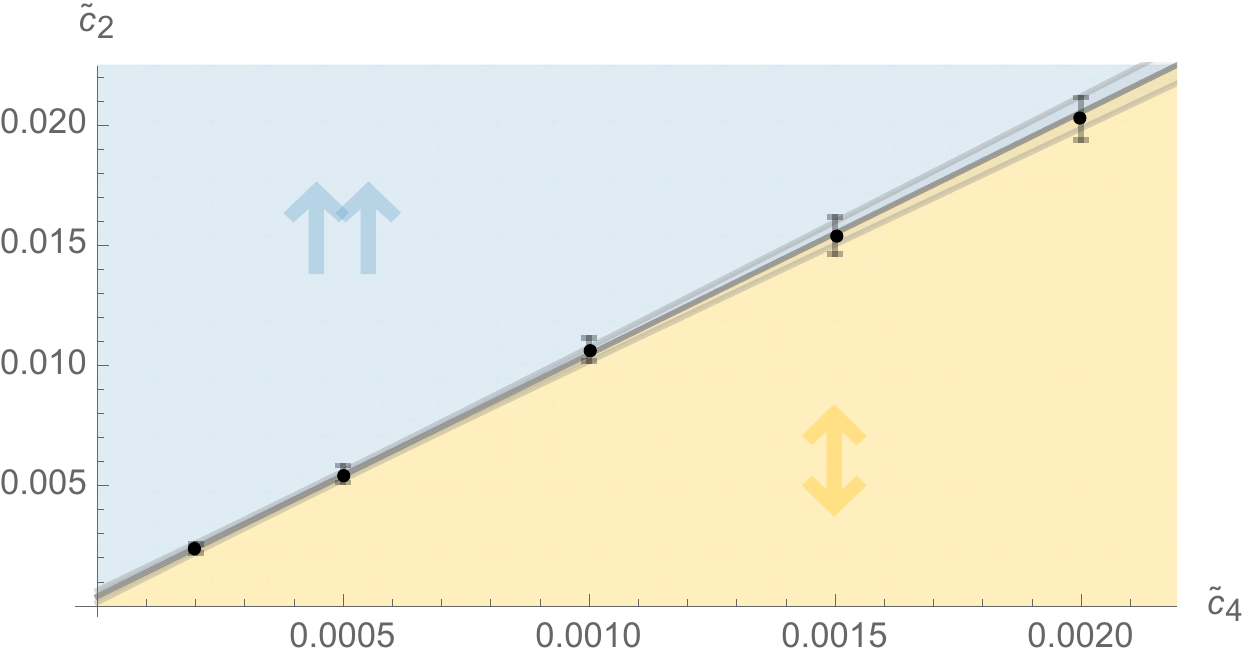}
\caption{Large $N$ extrapolation of the $\updownarrow \to \uparrow\uparrow$ transition line. Pale-gray stripes represent $68\%$ confidence intervals.}
\label{D2O}
\end{figure}
 
Based on the analysis of $a(N)$ and $b(N)$ from Figure \ref{AB}, the $\updownarrow \to \uparrow\uparrow$ transition line in the large $N$ limit extrapolates to
\begin{subequations}
\begin{align}
    C&: \quad \widetilde{c}_2 = - 0.03(7)\sqrt{\widetilde{c}_4} + 13(3)\widetilde{c}_4,
    \\
    \chi&: \quad \widetilde{c}_2 = + 0.01(1)\sqrt{\widetilde{c}_4} + 10.5(5)\widetilde{c}_4,
\end{align}
\label{CandChi}
\end{subequations}
These two expressions agree, as they should, or we could otherwise conclude that the triple point is located at the origin, and that 3-phase regime exists throughout the parameter space. Apparently, the $\sqrt{\widetilde{c}_4}$ effect completely disappears, since both square root coefficients are indistinguishable from zero. 

Equation of the $\updownarrow \to \uparrow\uparrow$ line in Figure \ref{D2O}, obtained from from linear fit through large $N$ limits at fixed $\widetilde{c}_4$, reads
\begin{equation}
     \chi: \quad \widetilde{c}_2 = + 0.0004(3) + 10.1(5)\widetilde{c}_4,
 \label{infD2O}
\end{equation}
which agrees with estimates in \eqref{CandChi} and Table \ref{comparison}.
Based on the extrapolation estimates of points that with increasing matrix size switch from 2-phase to 3-phase regime, there exists a possibility of systematic error from such still unidentified points, that could yield a lower true slope in \eqref{infD2O}. Namely, as triple point slides towards zero, it deforms the about-to-be-shortened end of the transition line close to it towards the less slanted $\updownarrow \to \uparrow\downarrow$ transition line. Also, inclusion of the $\widetilde{c}_4^{\,3/2}$ term into \eqref{c2(c4)} gives somewhat higher estimates for the linear term, although consistent with the reported ones.

The smallest $\widetilde{c}_4$ for which we detected change from 2-phase to 3-phase regime is  $\widetilde{c}_4 = 0.005$ at $N=28$. For all $\widetilde{c}_4 < 0.005$ and $N\leq50$ we see only two phases. This implies that $\updownarrow \to \uparrow\uparrow$ transition line ends in the triple point at $\widetilde{c}_4(T) \leq 0.005$.

\section{Conclusion}

We detailedly tested several choices for scaling of terms in the action of our model and chose the convergent albeit non-standard one: $\nu_2=1$, $\nu_4=1$, $\nu_r=0$, $\nu_k=0$. The choice replicated the known results for the PP model. Varying scalings around this choice led to transition lines without stable non-trivial infinite matrix limit. We semi-empirically determined equation \eqref{c2(c4)} of the $\updownarrow \to \uparrow\uparrow$ transition line when the kinetic term is turned on and found that it contains a part that captures the finite size effects and which disappears for larger matrices. The careful inspection of various scalings using two different approaches allowed us to non-trivially extrapolate this line from relatively small matrix sizes to the large $N$ limit.

We mapped phases of the model with turned off curvature in mass parameter-quatric coupling plane. The resulting diagram for $N=24$ is presented in Figure \ref{N24} and it consists, as expected, of three phases with different degree and kind of field eigenvalue activation. $\updownarrow \to \uparrow\downarrow$ and $\uparrow\downarrow \to \uparrow\uparrow$ transitions appear to be 3rd order. As for the $\updownarrow \to \uparrow\uparrow$ transition, specific heat for large matrices is practically constant compared to its large mass limit and fine details are buried under the data uncertainty. In contrast, peaks of susceptibility are well resolved and have a slight positive scaling with matrix size, and the transition appears to be of mixed 2nd and 3rd order and does not fall into the Ising universality class. It might well be that a mere presence of the matrix phase intermediary states near the $\updownarrow \to \uparrow\uparrow$ line interferes with the Ising-like behaviour. In the phase-coexistence region where the phases meet, 1st and 2nd order transitions are detected. This region shrinks with increasing matrix size, and it is expected to collapse into a triple point in the infinite limit. Another possible explanation for the non-Ising behaviour might be that the triple point actually lies at the very origin, and that what looks like the $\updownarrow \to \uparrow\uparrow$ border contains a $\uparrow\downarrow$ crack that reveals itself at larger matrix sizes.

In Figure \ref{D2O}, we extrapolated this border using matrices of sizes $N \leq 40$ and observed a convincing convergence. The extrapolated line radiates from the the origin with the slope $10.1(5)$. This could be the consequence of shortness of the $\updownarrow \to \uparrow\uparrow$ line, but clear disappearance of square root effects in Figure \ref{AB} indicates that the line is indeed linear. This linear behaviour is observed also on the fuzzy sphere \cite{triple}. We also demonstrated phase diagram convergence on token points from $\updownarrow \to \uparrow\downarrow$ and $\uparrow\downarrow \to \uparrow\uparrow$ transition lines. This is the part of the ongoing work of finding their large $N$ limits.

The triple point of the model is estimated to lie at $\widetilde{c}_4(T) \leq 0.005$. This is significantly smaller than the fuzzy sphere model value $\widetilde{c}_4(T) = 0.021(2)$ \cite{triple}, especially when larger matrices could pull it even closer to the origin. We still do not have enough data to extrapolate its final position. Once we find the limits of the remaining transition lines, we will be able to pinpoint it properly.

We also plan to compare these extrapolated lines with recent analytical results \cite{analytical1} for the fuzzy sphere in regimes where the two models could behave similarly, namely $\updownarrow \to \uparrow\downarrow$ line for large $\widetilde{c}_4$, where they should mimic the PP model, and $\uparrow\downarrow \to \uparrow\uparrow$ line where kinetic terms grow smaller as the field, up to a prefactor, oscillates closer to identity matrix.

While inspecting the scaling of the curvature term, we confirmed that it alters both eigenvalue distribution and the border of the $\uparrow\downarrow$ phase. Based on a cross section of the diagram, it seems that $\updownarrow \to \uparrow\downarrow$ line gets shifted proportionally to the curvature parameter $\widetilde{c}_r$ and to the scaled maximal eigenvalue of the curvature. The next important step is to see how it affects the full model in order to shed more light on its connection to renormalizability: the work on it is on the way.

\appendix
\section{Critical exponents and transition order}
\label{section:exponents}

We performed more detailed analysis of the large matrix transition limit at 3 points, corresponding to the clear two-phase regime ($\widetilde{c}_4=0.0001$), to the clear three-phase regime ($\widetilde{c}_4=1.0$) and to the phase coexistence regime near the triple point ($\widetilde{c}_4=0.01$).

To determine the universality class of our model's transitions we used the standard technique of finite size scaling. Mass parameter played the role of temperature and we defined reduced temperature $t$ near the critical $\widetilde{c}_2^{\,*}$ as
\begin{equation}
    t=1-\frac{\widetilde{c}_2}{\widetilde{c}_2^{\,*}}.
\end{equation}
In a nutshell, we consider the scalable part  $Q_s$ of quantity $Q$ to go as 
\begin{equation}
    Q_s(t)=N^{\epsilon_Q/\nu}\widetilde{Q}_s(tN^{1/\nu})
\end{equation}
near the transition, $\epsilon_Q$ being its critical exponent, and $\nu$ the critical exponent of correlation length. Unknown functions $\widetilde{Q}_s$ can be determined by varying $\widetilde{c}_2^{\,*}$, $\nu$ and exponents $\epsilon_Q$ until data for different $N$ collapse onto the same curve in some vicinity of the critical point. Also, if $Q$ peaks at the critical point, we can fit $Q_{\max} \sim N^{\epsilon_Q/\nu}$, while the position of the maximum $\widetilde{c}_2^{\,*}(N)$ approaches the true critical point as $\widetilde{c}_2^{\,*}(N)-\widetilde{c}_2^{\,*}\sim 1/N^{1/\nu}$. Following the convention, we denote the exponents of $C$, $M$ and $\chi$ as $\alpha$, $-\beta$ and $\gamma$ respectively. 

In \cite{hPT}, mixed order transitions are considered. They are classified as $(m,m')$ by lowest order derivatives of free energy with respect to temperature ($m$) and magnetic field ($m'$) that exhibit singular behaviour. In general, $m$ and $m'$ can differ. Let $A$ be a generalization of $\alpha$ -- the critical exponent of the lowest order temperature derivative of free energy that exhibits singular behaviour -- and similarly, $G$ generalization of $\gamma$. In a space of dimension $d$, $m=m'$ transition satisfies \cite{hPT}:
\begin{equation}
    (m-1)A + m\beta + G = m(m-1),
    \qquad
    \qquad
    m-A=\nu d.
\end{equation}
In the case of 2nd order transition, the first relation reduces to a familiar constraint
\begin{equation}
    \alpha + 2\beta + \gamma = 2.
    \label{constraint}
\end{equation}

\begin{figure}[t]
\centering
\includegraphics[scale=0.55]{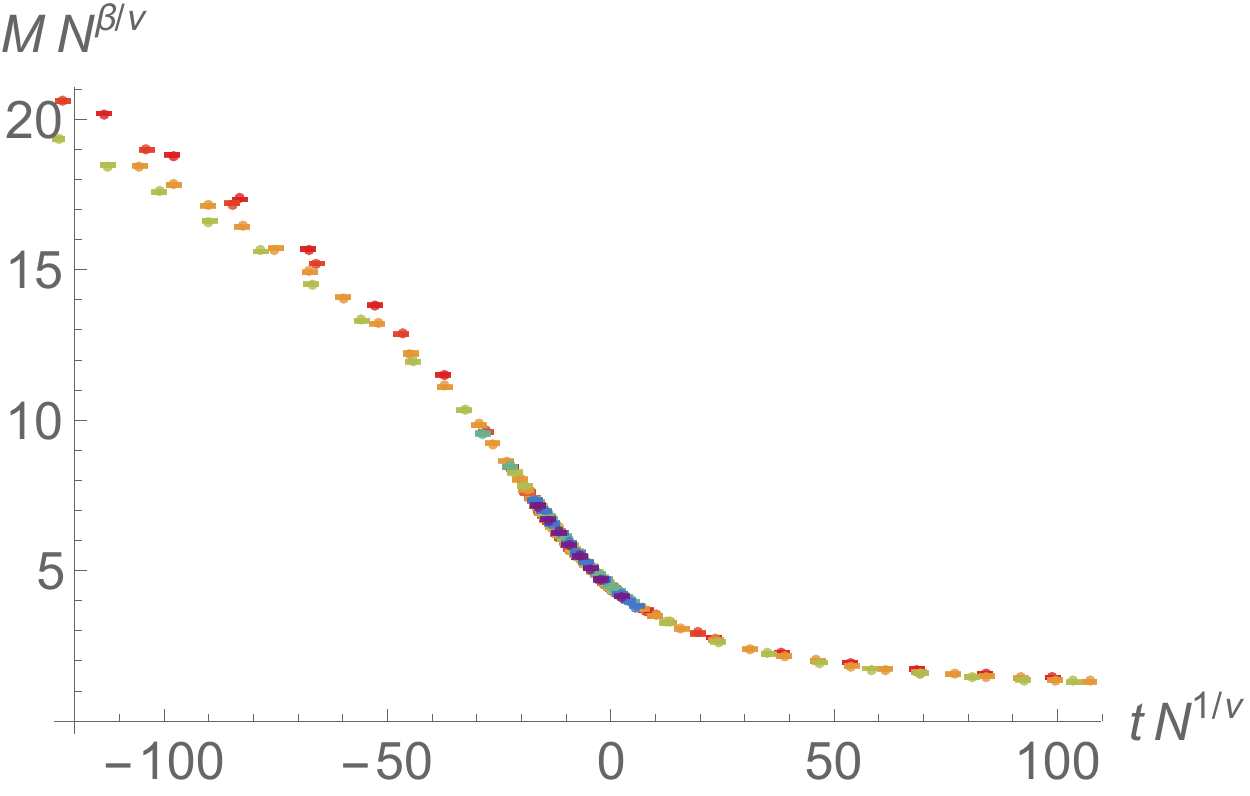}
\includegraphics[scale=0.55]{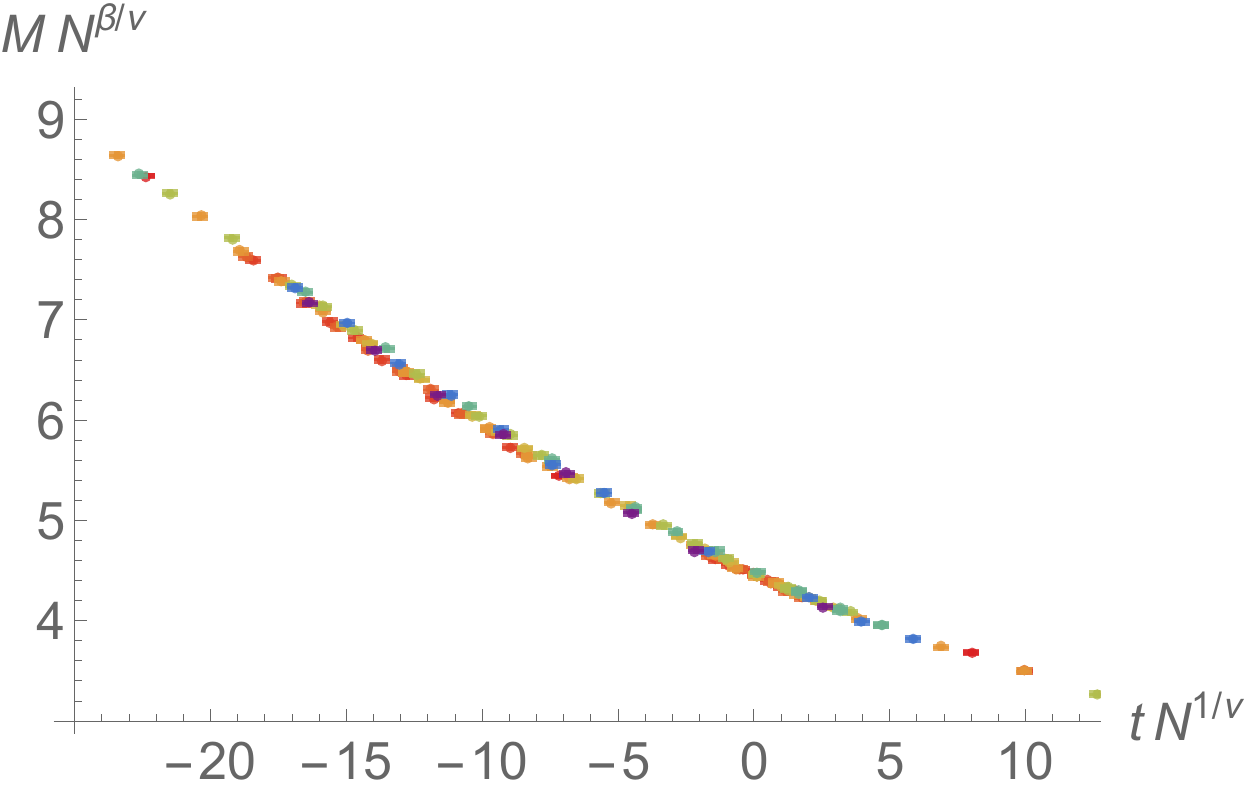}
\includegraphics[scale=0.55]{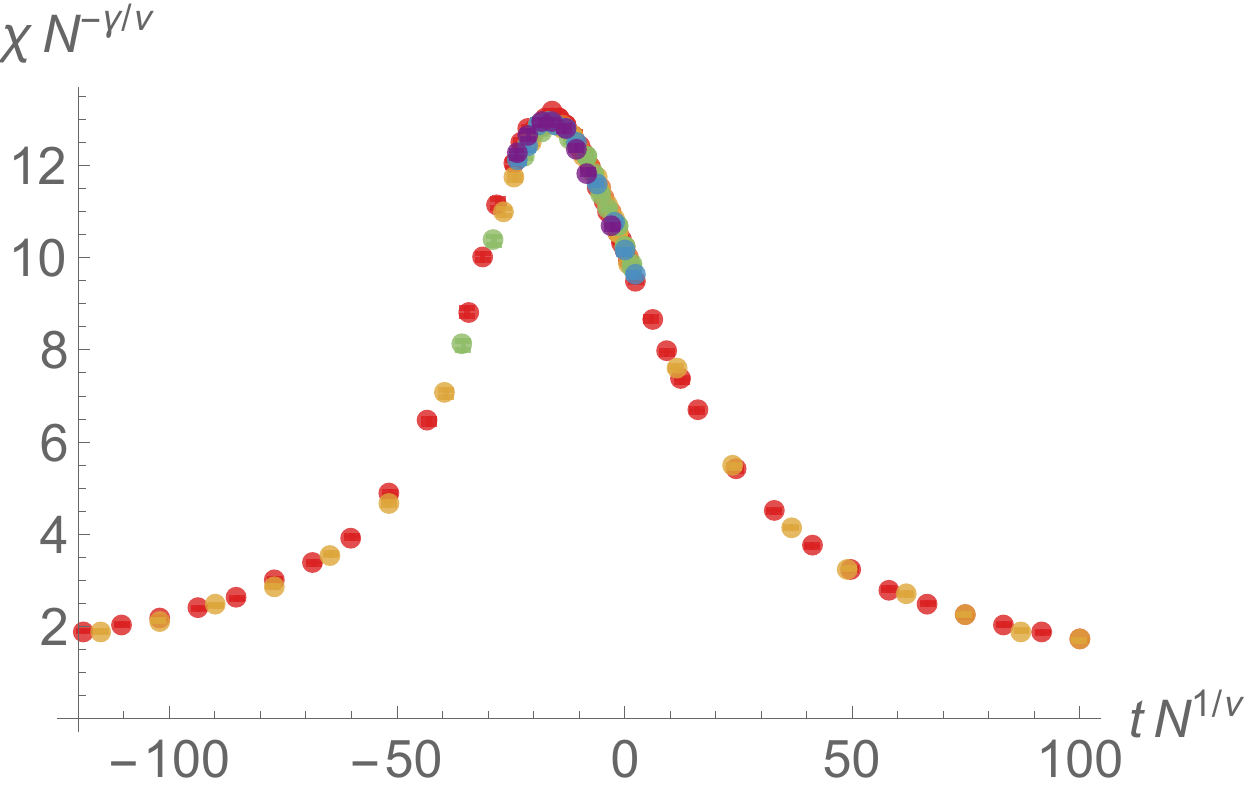}
\includegraphics[scale=0.55]{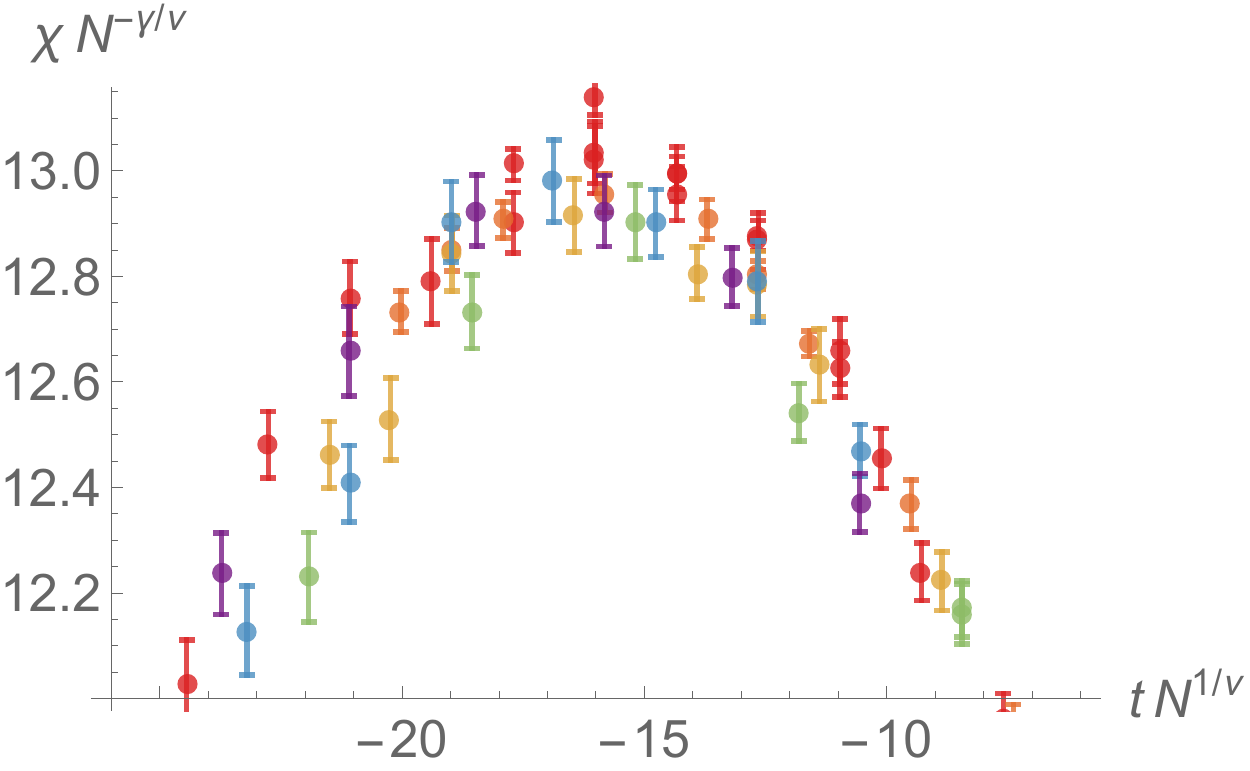}
\caption{Collapsed diagrams for $\updownarrow \to \uparrow\uparrow$ transition at $\widetilde{c}_4=0.0001$. Critical exponents are $\nu=1.00(2)$, $\beta=0.40(2)$ and  $\gamma=0.05(1)$. Different colors represent different matrix sizes up to $N=50$.}
\label{collapsed1}
\end{figure}

\begin{figure}[b]
\centering
\includegraphics[scale=0.55]{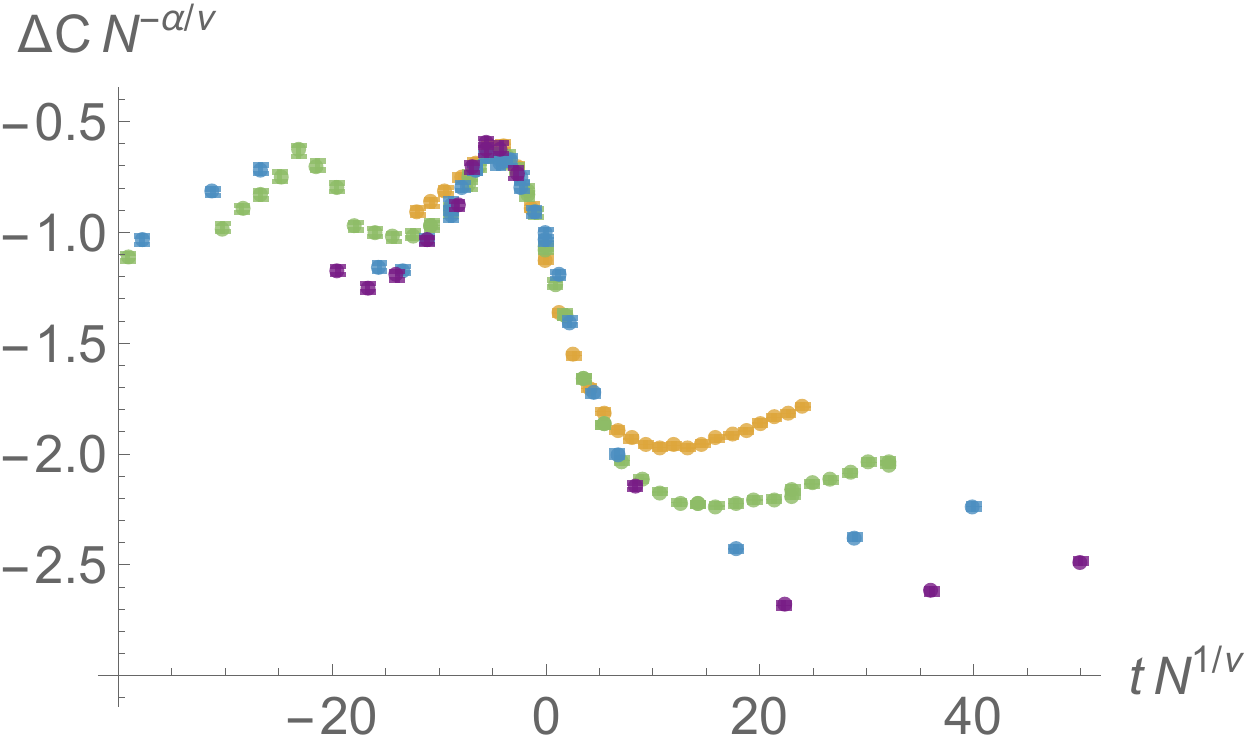}
\includegraphics[scale=0.55]{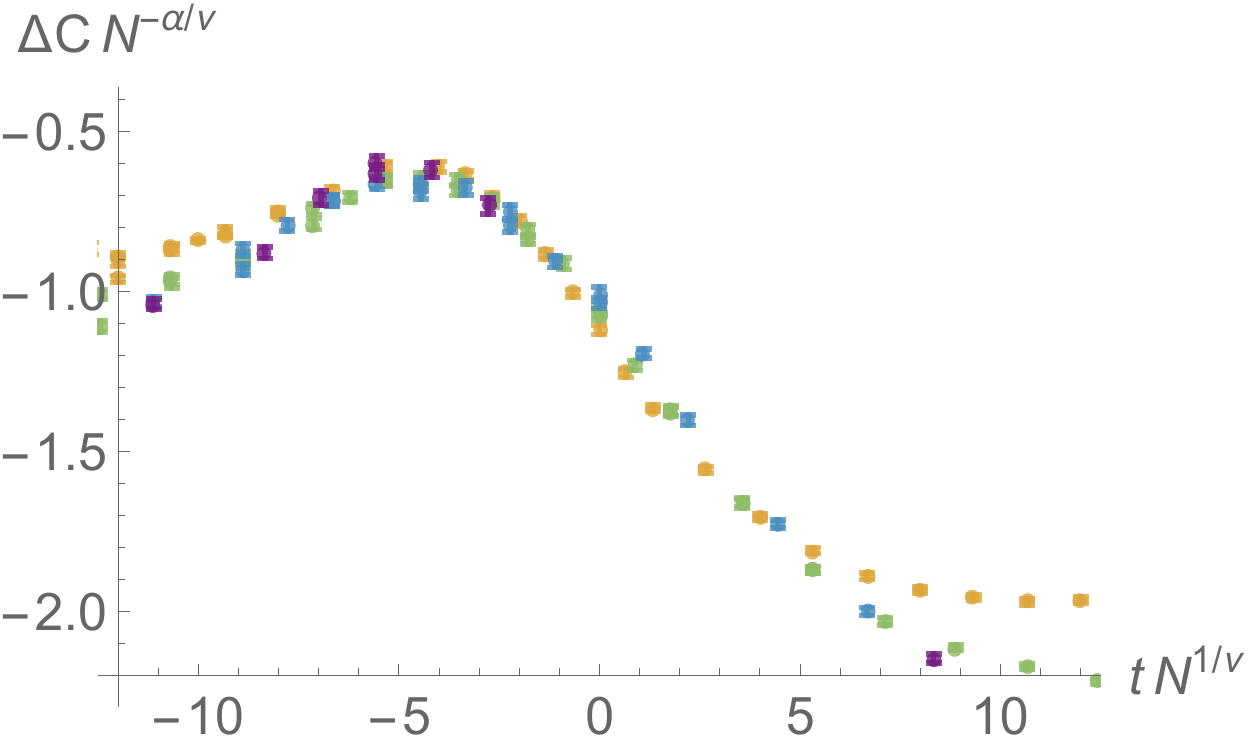}
\includegraphics[scale=0.53]{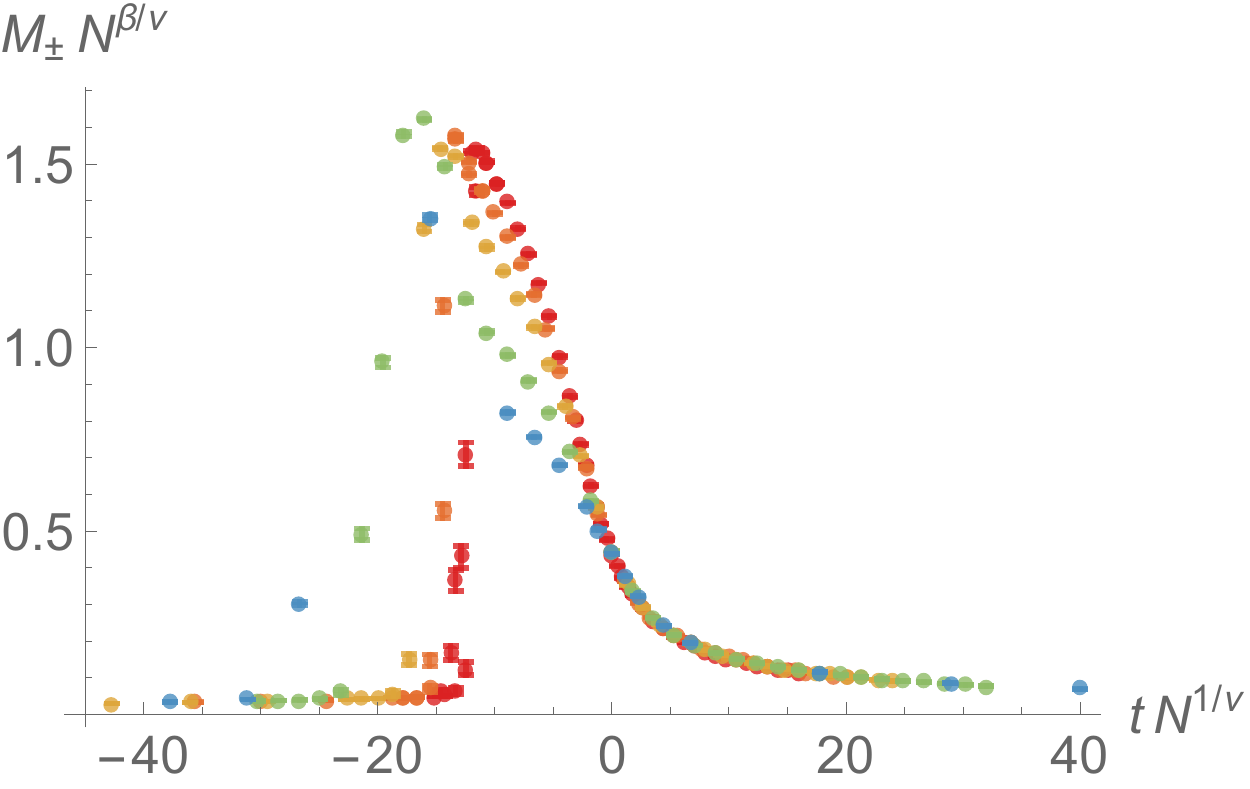}
\includegraphics[scale=0.53]{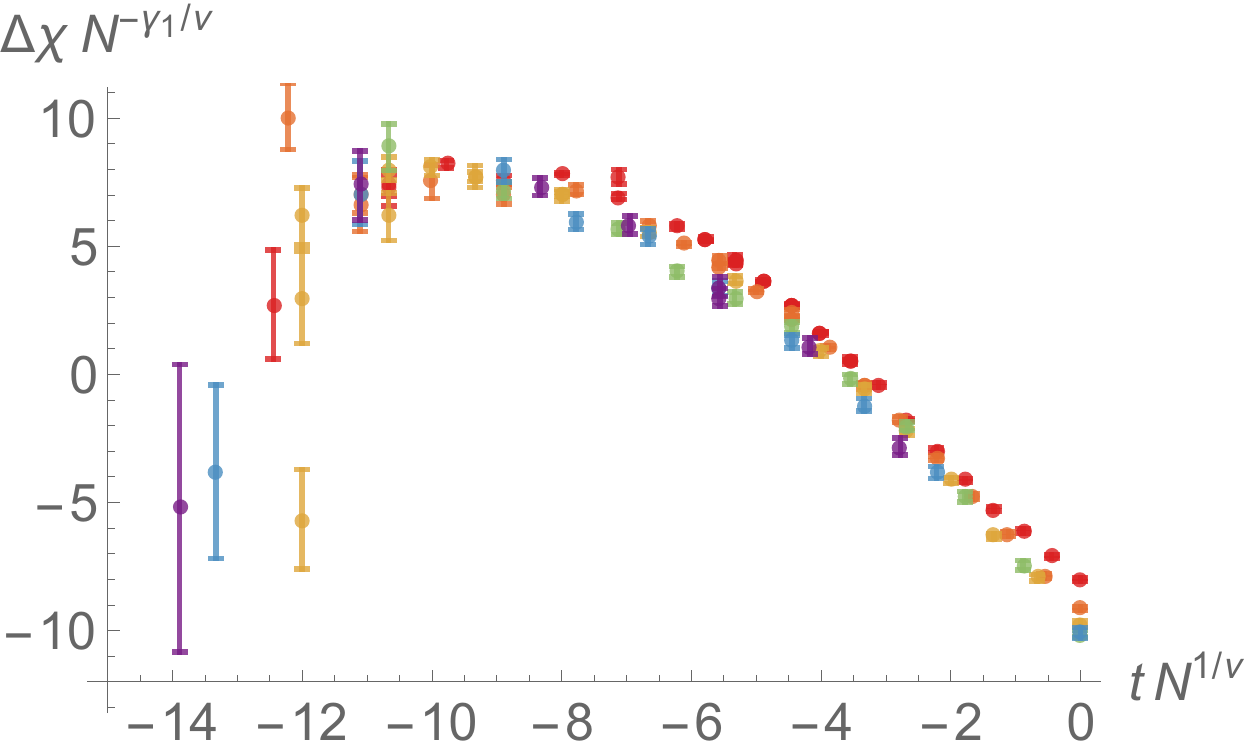}
\caption{Collapsed diagrams for $\updownarrow \to \uparrow\downarrow$ transition at $\widetilde{c}_4=1$. Critical exponents are $\nu=1.00(15)$, $\alpha=-0.41(6)$, $\beta=0.42(2)$ and  $\gamma=-0.99(7)$. $\Delta C=C-0.84(6)=-0.67(14)\cdot N^{\alpha/\nu}$ and $\gamma_1$ is the exponent of the correction to the scaling behaviour of susceptibility $\Delta\chi=\chi-1.13(2)\cdot N^{\gamma/\nu}=7.5(4)\cdot N^{-2.00(6)}$. Different colors represent different matrix sizes up to $N=50$.}
\label{collapsed2}
\end{figure}

\noindent The second relation implies that when there is a discontinuity in derivative ($A=0$), it must hold
\begin{equation}
    \nu = m/d.
\end{equation}

In Figure \ref{collapsed1}, we see collapsed data for $\updownarrow \to \uparrow\uparrow$ transition at $\widetilde{c}_4=0.0001$. One might expect it to belong to the Ising universality class, and indeed shapes of $\chi$ and $M$ look promising. However, their critical exponents differ as we can see in Table \ref{tab:exponents}. The transition appears to be weakly $(3,2)$ order, since $C$ remains finite and $\chi$ weakly diverges. Specific heat exhibits the familiar kink around its asymptotic value $0.5$. For larger matrices even this is hidden by errorbars and $C$ appears constant $C\approx 0.50(1)$. In the infinite limit it could develop discontinuity or a sharp edge, leading to either 2nd or 3rd order transition. That this transition cannot be 2nd order can be illustrated by analysing critical exponents. Even if we assume non-diverging $\alpha=0$ discontinuity in $C$ masked by errors, our exponents (Table \ref{tab:exponents}) cannot satisfy \eqref{constraint}, adding up to $0.85(3)$ instead of 2. However, a 3rd order transition could explain both this discrepancy and the value $\nu=1$, if we assume that transition sees the compactified 3rd dimension of the $\mathfrak{h}^\text{tr}$ space:
\begin{equation}
    \nu = m/d = 3/3 =1.
\end{equation}

In Figure \ref{collapsed2}, we see collapsed data for $\updownarrow \to \uparrow\downarrow$ transition at $\widetilde{c}_4=1$. Both $C$ and $\chi$ remain finite, and the transition governed by the split in eigenvalue distribution is 3rd order, the same type as in the PP model. The $\uparrow\downarrow \to \uparrow\uparrow$ transition at this $\widetilde{c}_4$ shows nearly identical peak in  $C$ as $\updownarrow \to \uparrow\downarrow$ transition (nicely seen in green data of the top left plot in Figure \ref{collapsed2}) and it also appears to be 3rd order.

Near triple point, at $\widetilde{c}_4=0.01$, $\updownarrow$$+$$\uparrow\downarrow \to \uparrow\uparrow$ transition is 1st order and both $C$ and $\chi$ diverge with $\alpha/\nu=3.07(3)$, $\gamma/\nu=3.47(8)$.

\begin{table}[t]
\centering
\setlength{\tabcolsep}{1em}
\renewcommand{\arraystretch}{1.2}
\begin{tabular}{|c||c|c|c|c|}
    \hline
    model & $\alpha$ & $\beta$ & $\gamma$ & $\nu$ 
    \\
    \hline
   $\updownarrow \to \uparrow\uparrow$ @ $\widetilde{c}_4=0.0001$  & $\leq 0$ & $0.40(2)$ & $0.05(1)$  & $1.00(2)$
    \\
    $\updownarrow \to \uparrow\downarrow$ @ $\widetilde{c}_4=1.0000$ & $-0.41(6)$ & $0.42(2)$ & $-0.99(7)$ & $1.00(15)$
     \\
    Ising $2D$ & $0 \, (\log)$ & $1/8$ & $7/4$ & $1$
    \\
    Ising $3D$ & $0.110(1)$ & $0.3265(3)$ & $1.2372(5)$ & $0.6301(4)$
    \\
    \hline
\end{tabular}
\vspace{10pt}
\caption{\label{tab:exponents} Comparison of critical exponents of our model and the Ising model \cite{exponents}.}
\end{table}

We have detected both 1st and 2nd order transitions for different matrix sizes in different parts of parameter space. For small $\widetilde{c}_4$ we have well separated $\updownarrow$ and  $\uparrow\uparrow$ phases. For large $\widetilde{c}_4$ all 3 phases are well separated. For the intermediary values of $\widetilde{c}_4$ we encounter phase coexistence region that grows smaller with increasing matrix size and hopefully collapses into a triple point in the infinite limit. In that region smaller $\widetilde{c}_4$ show $\updownarrow$$+$$\uparrow\downarrow$ mixture of phases, while larger $\widetilde{c}_4$ show $\uparrow\downarrow'$$+$$\uparrow\downarrow''$ mixture of phases (bottom center plot in Figure \ref{thermo}). The former is more symmetric and apparently produces 2nd order transitions, while latter is less symmetric and leads to 1st order transitions. A heuristic behind it is: a pile of needles is almost as smooth as a ball, but three needles will prick.

\section{Curvature term}
\label{section:curvature}

We wish to briefly examine the scaling of the curvature term of $S_{\text{R}+\text{PP}}$ by looking at its effects on top of the PP model, without complications of the kinetic term. We will consider the relevant case where $c_r>0$. 

The NC curvature of the model is a negative diagonal matrix 
\begin{equation}
    R_{jj} = \mathcal{R} + 8 - 
    \begin{cases}
    16j, & 1<j<N,
    \\
    8N, & j=N,
    \end{cases}
\end{equation}
where $\mathcal{R} = 15/2$; in simulation we erroneously used $\mathcal{R} = 15/4$ but that does not change the conclusions of this section because they depend on the $\mathcal{O}(N)$ part of the curvature.  Diagonality yields $\Tr\,(R\Phi^2)=R_{jj}(\Phi^2)_{jj}$, bounding the curvature term in the action by
\begin{equation}
\Tr\left(
c_r\min_j\abs{R_{jj}}\Phi^2 
\right)
\le
\Tr\left(
c_r \abs{R}\Phi^2
\right)
\le
\Tr\left(
c_r\max_j\abs{R_{jj}}\Phi^2
\right),
\end{equation}
which translates to
\begin{equation}
\Tr\left(
(8-\mathcal{R})\widetilde{c}_r\Phi^2 
\right)
\le
\Tr\left(
c_r \abs{R}\Phi^2
\right)
\le
\Tr\left(
\left(16N-(24+\mathcal{R})\right)\widetilde{c}_r\Phi^2
\right).
\end{equation}
Treating this as a bounded contribution to the mass term, we could naively expect it to be reflected in a deformation of the transition line $\widetilde{c}_2\to\widetilde{c}_{2,r}$
\begin{equation}
\widetilde{c}_2+\frac{8-\mathcal{R}}{N}\widetilde{c}_r
 \le \widetilde{c}_{2,r} \le
\widetilde{c}_2+\left(16-\frac{24+\mathcal{R}}{N}\right)\widetilde{c}_r.
\end{equation}
The wrong choice of scaling would change this into
\begin{equation}
 \widetilde{c}_2+\frac{8-\mathcal{R}}{N}\widetilde{c}_rN^{\Delta\nu_r}
 \le \widetilde{c}_{2,r} 
 \le \widetilde{c}_2+\left(16-\frac{24+\mathcal{R}}{N}\right)\widetilde{c}_rN^{\Delta\nu_r}.
 \label{boundR}
\end{equation}
This means that for $\Delta\nu_r<0$ we would practically see the PP case and for $\Delta\nu_r>0$ the $N^{\Delta\nu_r}$ runaway effect towards large values of the mass parameter.

\begin{figure}[t]
\centering
\includegraphics[scale=0.90]{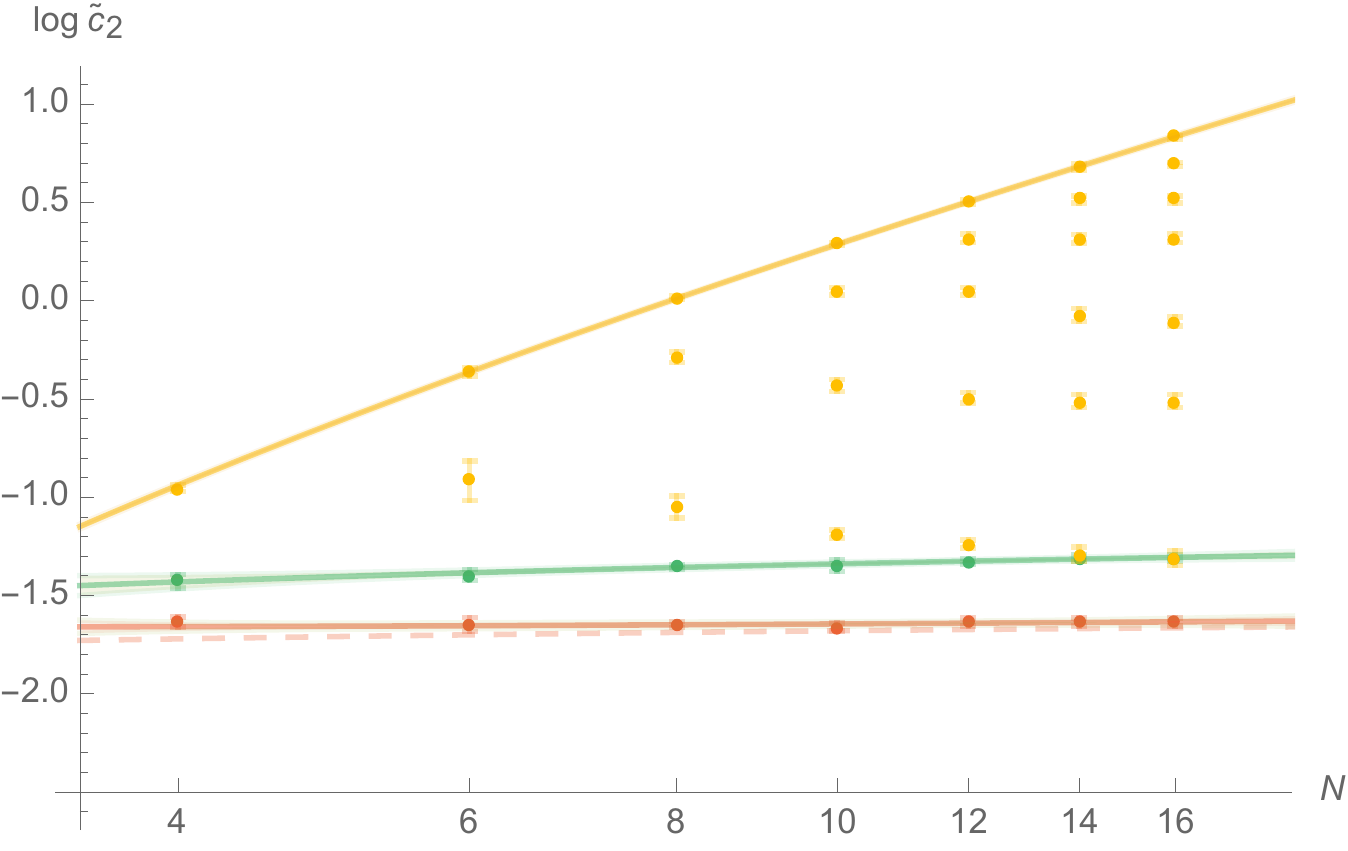}
\caption{$\updownarrow \to \uparrow\downarrow$ transition in the PP model with curvature for $\widetilde{c}_4=0.01$, $\widetilde{c}_r=0.01$, $4\leq N \leq 16$ and fixed $\nu_2=\nu_4=1$, observed as peaks in $\chi$ and $M$. The green/center data represents the correct choice of scaling $\nu_r=0$, the orange/top $\Delta\nu_r=+1$, the red/bottom  $\Delta\nu_r=-1$ and the pale-red dashed line the PP model. For $\Delta\nu_r=+1$ and fixed $N$, magnetization peaks $N/2-1$ times (unconnected orange dots) with increasing $\widetilde{c}_2$ until $\chi$ reaches its maximum and eigenvalue distribution splits in two, causing the phase transition (connected orange dots). Errorbars are mostly covered by data markers and pale coloured stripes represent the $68\%$ confidence intervals.}
\label{scalingR}
\end{figure}

This is exactly what happens in Figure \ref{scalingR} to the slanted orange line 
\begin{equation}
 1.01(3)\log N - 1.83(9) - \frac{2.0(2)}{N},
\end{equation}
which fits very well with the expansion of the left-hand side of \eqref{boundR} (with \eqref{c2(c4)PP} substituted)
\begin{equation}
 \log N - 1.83 - \frac{2.98}{N},
\end{equation}
and its slope $1.01(3)$ with $\Delta\nu_r=1$.

There are multiple peaks of $M$ for $\Delta\nu_r=1$, fixed $N$ and varying $\widetilde{c}_2$, marked by unconnected orange dots in Figure \ref{scalingR}, complicating identification of the phase transition. However, only the topmost of them coincide with the peaks of $\chi$ which we use as the indicator of the phase transition. 

\begin{figure}[b]
\vspace{5pt}
\centering
\includegraphics[scale=0.45]{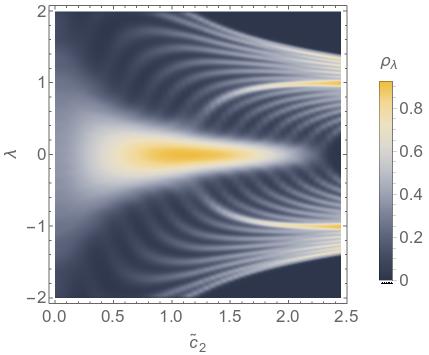}
\hspace{10pt}
\includegraphics[scale=0.47]{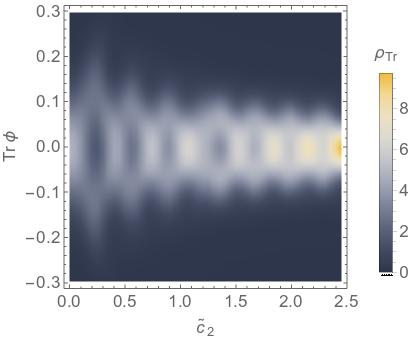}
\caption{Eigenvalue (left) and trace (right) distributions of the PP model with $\Delta\nu_r=+1$ curvature for $N=16$, $\widetilde{c}_4=0.01$, $\widetilde{c}_r=0.01$, $\nu_2=\nu_4=1$ and varying values of $\widetilde{c}_2$. Brighter regions correspond to larger values and peaks of distributions. Central bright region in the left plot, that dims and widens to the left, depicts the $\updownarrow$ phase which at around $\widetilde{c}_2\approx2.3$ completely splits into two cuts of the $\uparrow\downarrow$ phase. Two thicker bright lines in the eigenvalue distribution plot are due to degenerate eigenvalues of the curvature matrix. Eigenvalues are expressed in units of $\sqrt{\Tr\Phi^2/N}$ and traces in units of $\sqrt{N\Tr\Phi^2}$.}
\label{eigenR}
\end{figure}

This is further confirmed by inspection of the eigenvalue distribution in Figure \ref{eigenR}. As the mass parameter increases, one by one separate peaks break off the edge of the shrinking deformed Wigner semi-circle. Meanwhile the trace distribution stays centered at zero. We interpret this as curvature eigenvalues activating one by one with alternating signs, causing the magnetization to fluctuate and form peaks, and trace distribution to expand and contract around zero mean. This continues until all eigenvalues separate from the bulk, susceptibility peaks and system transitions into a modified matrix phase around $\widetilde{c}_2\approx2.3$.

The left-hand side of \eqref{boundR} also predicts the shift between the $\Delta\nu_r=0$ and the PP-line to be less than $16\,\widetilde{c}_r=0.16$ and the actual difference at $N=16$ is $0.15(4)$. As for the $\Delta\nu_r=-1$ line, it is practically indiscernible from the PP-line, as expected.

\section{Transition line coefficients}
\label{section:coefAB}

\begin{table}[b]
\centering
\setlength{\tabcolsep}{1em}
\renewcommand{\arraystretch}{1.2}
\begin{tabular}{|c|c|c|crr|}
    \hline
    $\mathcal{O}$ & $\Delta\nu_k$ & $\widetilde{c}_4$ &  expression & method I &  method II 
    \\
    \hline
    $N$  & $-1.0\phantom{-}$ & $1\cdot10^{-5}$ & $b_0$ & $11.4(9)$ & $10.5(5)$
    \\
    \hline
    \multirow{2}{*}{$\sqrt{N}$}  & $-1.0\phantom{-}$ & $1\cdot10^{-5}$ & $b_1$ & $-36(6)$ & $-31(4)$
    \\
    & $-0.5\phantom{-}$ & $5\cdot10^{-3}$ & $b_0$ & $10.6(6)$ & $10.5(5)$
    \\
    \hline
    \multirow{4}{*}{$\displaystyle 1$}  & $-1.0\phantom{-}$ & $1\cdot10^{-5}$ & $b_2+a_0/\sqrt{\widetilde{c}_4}$ & $55(9)$ & $46(9)$
    \\
    & $-0.5\phantom{-}$ & $5\cdot10^{-3}$ & $b_1+a_0/\sqrt{\widetilde{c}_4}$ & $-32(3)$ & $-31(4)$
    \\
    & $0.0$ & $1\cdot10^{-3}$ & $b_0+a_0/\sqrt{\widetilde{c}_4}$ & $10.7(5)$ & $10.8(5)$
    \\
    & $2.0$ & $1\cdot10^{-2}$ & $a_0$ & $-0.00(4)$ & $0.1(2)$
    \\
    \hline
    \multirow{4}{*}{$\displaystyle\frac{1}{\sqrt{N}}$}
    & $-0.5\phantom{-}$  & $5\cdot10^{-3}$ & $b_2+a_1/\sqrt{\widetilde{c}_4}$  & $50(3)$ & $44(9)$
    \\
    & $0.0$ & $1\cdot10^{-3}$ & $b_1+a_1/\sqrt{\widetilde{c}_4}$ & $-24(4)$ & $-29(4)$
    \\
    & $0.5$ & \multirow{2}{*}{$1\cdot10^{-2}$} & $b_0+a_1/\sqrt{\widetilde{c}_4}$ & $12(1)$ & $11.2(9)$
    \\    
    & $2.0$ & & $a_1$ & $-0.0(2)$ & $0.07(7)$
    \\
    \hline
    \multirow{5}{*}{$\displaystyle\frac{1}{N}$} & $-1.0\phantom{-}$ & $1\cdot10^{-5}$ & $a_2+\beta_4\sqrt{\widetilde{c}_4}\phantom{+}$ & $1.99(7)$ & $2.0(2)$
    \\
    & $0.0$ & $1\cdot10^{-3}$ & $b_2+a_2/\sqrt{\widetilde{c}_4}$ & $86(6)$ & $109(9)$
    \\
    & $0.5$ & \multirow{3}{*}{$1\cdot10^{-2}$} & $b_1+a_2/\sqrt{\widetilde{c}_4}$ & $-17(8)$ & $-11(4)$
    \\
    & $1.0$ & & $a_2+b_0\sqrt{\widetilde{c}_4}\phantom{+}$ & $3.3(3)$ & $3.1(1)$
    \\
    & $2.0$ & & $a_2$ & $2.3(3)$ & $2.06(9)$
    \\
    \hline
\end{tabular}
\vspace{10pt}
\caption{\label{tab:AB} Comparison of the estimates of $a_i$ and $b_i$ using fits for different $\Delta\nu_k$ and fixed $\widetilde{c}_4$ (method I, using \eqref{c2(N)}) to the estimates from $\nu_k=0$ and variable $\widetilde{c}_4$ and $N$ (method II, using \eqref{Puiseux}).}
\label{comparison}
\end{table}

In order to access the large $N$ convergence of the $\updownarrow \to \uparrow\uparrow$ transition line and subsequently that of $a(N)$ and $b(N)$, we compared two approaches: 
\begin{itemize}
     \item method I: for fixed $\widetilde{c}_4$ and various fixed $\nu_k$, we varied $N$ and for each detected $\widetilde{c}_2(N)$ at which transition occurs; we then fitted the $1/\sqrt{N}$-expansion of \eqref{c2(N)} to get the combinations of  $a_i$, $b_i$ and $\widetilde{c}_4$  (Table \ref{tab:AB});
     \item method II: for fixed $N$ and $\nu_k=0$, we constructed the transition line for a range of $\widetilde{c}_4$ and then extracted $a(N)$ and $b(N)$ using \eqref{c2(c4)}; we then varied $N$ and fitted series \eqref{Puiseux} to get $a_i$ and $b_i$ (Figure \ref{AB}).
\end{itemize}

Applying the method II to the $\chi$-data from Figure \ref{AB}, we got the expansions \eqref{Puiseux}, where we used the lowest order polynomial in $1/\sqrt{N}$ that fits well with the data. The higher terms turn out to be indiscernible form zero within their large uncertainties. The $C$-data have much less predictive power since the peaks of $C$ are wide, skewed, nearly flat and do not scale with $N$, unlike the peaks in $\chi$ which are well resolved. 

The comparison of these two approaches is given in Table \ref{tab:AB}: we see that the choice of $\nu_k=0$ scaling of the kinetic term leads to consistent values for coefficients of the transition line. Also, with increasing matrix size $\Delta\nu_k>0$ transition points collapse to zero in the predicted manner which is for $\Delta\nu_k\geq1$ practically linear.

\acknowledgments

This work was supported by the Serbian Ministry of Education, Science and Technological Development Grant
ON171031 and by COST Action MP1405. We would like to thank Prof Maja Buri\'{c}, Prof Denjoe O'Connor and Dr Samuel Kov\'{a}\v{c}ik for valuable discussions and DIAS for hospitality and financial support.


\begin{thebibliography}{99}

\bibitem{NC1}
    H. S. Snyder, 
    \emph{Quantized Spacetime}.
    \href{https://doi.org/10.1103/PhysRev.71.38}{\emph{Phys. Rev.} {\bf 71} (1947) 38}
    [\href{https://inspirehep.net/search?p=find+10.1103/PhysRev.71.38}{\inspire}]

\bibitem{NC2} 
    J. Bellissard, A. van Elst, H. Schulz-Baldes, 
    \emph{The noncommutative geometry of the quantum Hall effect},
    \href{https://doi.org/10.1063/1.530758}{\emph{JMP} {\bf 35} (1994) 5373}
    [\href{https://arxiv.org/abs/cond-mat/9411052}{\tt arXiv:cond-mat/9411052}]

\bibitem{NC3} 
    K. Fujii, 
    \emph{From quantum optics to non-commutative geometry: A Non-commutative version of the Hopf bundle, Veronese mapping and spin representation},
    [\href{https://arxiv.org/abs/quant-ph/0502174}{\tt arXiv:quant-ph/0502174}]
    [\href{https://inspirehep.net/search?p=find+EPRINT+quant-ph/0502174}{\inspire}]

\bibitem{string1}
    N. Seiberg, E. Witten, 
    \emph{String theory and noncommutative geometry},
    \href{https://doi.org/10.1088/1126-6708/1999/09/032}{\emph{JHEP} {\bf 09} (1999) 032}
    [\href{https://arxiv.org/abs/hep-th/9908142}{\tt arXiv:hep-th/9908142}]
    [\href{https://inspirehep.net/search?p=find+EPRINT+hep-th/9908142}{\inspire}]

\bibitem{string2}
    L. Freidel, R. G. Leighb, \DJ. Mini\'{c},
    \emph{Intrinsic non-commutativity of closed string theory},
    \href{https://doi.org/10.1007/JHEP09(2017)060}{\emph{JHEP} {\bf 09} (2017) 060}
    [\href{https://arxiv.org/abs/1706.03305}{\tt arXiv:1706.03305 [hep-th]}]
    [\href{https://inspirehep.net/search?p=find+EPRINT+arXiv:1706.03305}{\inspire}]
    	
\bibitem{UV/IR1}
    S. Minwalla, M. Van Raamsdonk, 
    \emph{Noncommutative perturbative dynamics}, 
    \href{https://doi.org/10.1088/1126-6708/2000/02/020}{\emph{JHEP} {\bf 02} (2000) 020}
    [\href{https://arxiv.org/abs/hep-th/9912072}{\tt arXiv:hep-th/9912072}]
    [\href{https://inspirehep.net/search?p=find+EPRINT+hep-th/9912072}{\inspire}]

\bibitem{UV/IR2}
    C. S. Chu, J. Madore, H. Steinacker, 
    \emph{Scaling limits of the fuzzy sphere at one loop}, 
    \href{https://doi.org/10.1088/1126-6708/2001/08/038}{\emph{JHEP} {\bf 08} (2001) 038} 
    [\href{https://arxiv.org/abs/hep-th/0106205}{\tt arXiv:hep-th/0106205}]
    [\href{https://inspirehep.net/search?p=find+EPRINT+hep-th/0106205}{\inspire}]

\bibitem{UV/IR3}
    B. P. Dolan, D. O'Connor, P. Pre\v{s}najder, 
    \emph{Matrix $\phi^4$ models on the fuzzy sphere and their continuum limits}, 
    \href{https://doi.org/10.1088/1126-6708/2002/03/013}{\emph{JHEP} {\bf 03} (2002) 013}
    [\href{https://arxiv.org/abs/hep-th/0109084}{\tt arXiv:hep-th/0109084}]
    [\href{https://inspirehep.net/search?p=find+EPRINT+hep-th/0109084}{\inspire}]

\bibitem{GW1} 
    H. Grosse, R. Wulkenhaar, 
    \emph{Renormalization
    of $\phi^{4}$-theory on noncommutative $\mathbb{R}^{2}$ in the matrix base}, 
    \href{https:///doi.org/10.1088/1126-6708/2003/12/019}{\emph{JHEP} {\bf 12} (2003) 019} 
    [\href{https://arxiv.org/abs/hep-th/0307017}{\tt arXiv:hep-th/0307017}]
    [\href{https://inspirehep.net/search?p=find+EPRINT+hep-th/0307017}{\inspire}]

\bibitem{GW2} 
    H. Grosse, R. Wulkenhaar, 
    \emph{Renormalization of $\phi^{4}$-theory on noncommutative $\mathbb{R}^{4}$ in the matrix base}, 
    \href{https:///doi.org/10.1007/s00220-004-1285-2}{\emph{Commun. Math. Phys.} {\bf 256} (2005) 305} 
    [\href{https://arxiv.org/abs/hep-th/0401128}{\tt arXiv:hep-th/0401128}]
    [\href{https://inspirehep.net/search?p=find+EPRINT+hep-th/0401128}{\inspire}]

\bibitem{GW3} 
    H. Grosse, R. Wulkenhaar, 
    \emph{The $\beta$-function in duality-covariant noncommutative $\phi^{4}$-theory}, \href{https://doi.org/10.1140/epjc/s2004-01853-x}{\emph{EPJ} {\bf C 35} (2004) 277} 
    [\href{https://arxiv.org/abs/hep-th/0402093}{\tt arXiv:hep-th/0402093}]
    [\href{https://inspirehep.net/search?p=find+EPRINT+hep-th/0402093}{\inspire}]

\bibitem{GW4} 
    M. Disertori, R. Gurau, J. Magnen, V. Rivasseau, 
    \emph{Vanishing of beta function of non commutative $\Phi_{4}^{4}$ theory to all orders},
    \href{https://doi.org/10.1016/j.physletb.2007.04.007}{\emph{Phys. Lett.} {\bf B 649} (2007) 95} 
    [\href{https://arxiv.org/abs/hep-th/0612251}{\tt arXiv:hep-th/0612251}]
    [\href{https://inspirehep.net/search?p=find+EPRINT+hep-th/0612251}{\inspire}]
    
\bibitem{GW5} 
    Z. Wang, 
    \emph{Constructive Renormalization of the $2$-Dimensional Grosse-Wulkenhaar Model}, 
    \href{https://doi.org/10.1007/s00023-018-0688-0}{\emph{Ann. Henri Poincar\'{e}} {\bf 19} (2018) 2435}
    [\href{https://arxiv.org/abs/1805.06365}{\tt arXiv:1805.06365 [math-ph]}]
    [\href{https://inspirehep.net/search?p=find+EPRINT+arXiv:1805.06365}{\inspire}]

\bibitem{rNC1}    
    M. Buri\'{c}, J. Madore, L. Nenadovi\'{c}, 
    \emph{Spinors on a curved noncommutative space: coupling to torsion and the Gross-Neveu model},
    \href{https://doi.org/10.1088/0264-9381/32/18/185018}{\emph{Class. Quant. Grav.} {\bf 32} (2015) 185}
    [\href{https://arxiv.org/abs/1502.00761}{\tt arXiv:1502.00761 [hep-th]}]
    [\href{https://inspirehep.net/search?p=find+EPRINT+arXiv:1502.00761}{\inspire}]
    
\bibitem{rNC2}    
    F. Vignes-Tourneret, 
    \emph{Renormalization of the Orientable Non-commutative Gross-Neveu Model},
    \href{https://doi.org/10.1007/s00023-006-0312-6}{\emph{Ann. Henri Poincar\'{e}} {\bf 8} (2007) 427}
    [\href{https://arxiv.org/abs/math-ph/0606069}{\tt arXiv:math-ph/0606069}]
    [\href{https://inspirehep.net/search?p=find+EPRINT+math-ph/0606069}{\inspire}]

\bibitem{rNC3}    
   D. J. Gross, A. Neveu, 
   \emph{Dynamical symmetry breaking in asymptotically free field theories},
   \href{https://doi.org/10.1103/PhysRevD.10.3235}{\emph{Phys. Rev.} {\bf D 10} (1974) 3235}
   [\href{https://inspirehep.net/search?p=find+10.1103/PhysRevD.10.3235}{\inspire}]

\bibitem{tHA}
    M. Buri\'{c}, M. Wohlgenannt, 
    \emph{Geometry of the Grosse-Wulkenhaar model}, 
    \href{https://doi.org/10.1007/JHEP03(2010)053}{\emph{JHEP} {\bf 03} (2010) 053}
    [\href{https://arxiv.org/abs/0902.3408}{\tt arXiv:0902.3408 [hep-th]}]
    [\href{https://inspirehep.net/search?p=find+EPRINT+arXiv:0902.3408}{\inspire}]

\bibitem{striped1}
    G. S. Gubser, S. L. Sondhi, 
    \emph{Phase structure of non-commutative scalar field theories}, 
    \href{https://doi.org/10.1016/S0550-3213(01)00108-0}{\emph{Nucl. Phys.} {\bf B 605} (2001) 395}
    [\href{https://arxiv.org/abs/hep-th/0006119}{\tt arXiv:hep-th/0006119}]
    [\href{https://inspirehep.net/search?p=find+EPRINT+hep-th/0006119}{\inspire}]
    
\bibitem{striped2}
    P. Castorina and D. Zappal\`{a},
    \emph{Spontaneous breaking of translational invariance in non-commutative $\lambda\phi^4$ theory in two dimensions},
    \href{https://doi.org/10.1103/PhysRevD.77.027703}{\emph{Phys. Rev.} {\bf D 77} (2008) 027703}
    [\href{https://arxiv.org/abs/0711.2659}{\tt arXiv:0711.2659 [hep-th]}]
    [\href{https://inspirehep.net/search?p=find+EPRINT+arXiv:0711.2659}{\inspire}]
    
\bibitem{striped3}
    H. Mej\'{i}a-D\'{i}az, W. Bietenholz, M. Panero,
    \emph{The continuum phase diagram of the $2d$ non-commutative $\lambda\phi^4$ model},
    \href{https://doi.org/10.1007/JHEP10(2014)056}{\emph{JHEP} {\bf 10} (2014) 056}
    [\href{https://arxiv.org/abs/1403.3318}{\tt arXiv:1403.3318 [hep-lat]}]
    [\href{https://inspirehep.net/search?p=find+EPRINT+arXiv:1403.3318}{\inspire}]
    
\bibitem{MFT} 
    B. Ydri, 
    \emph{Lectures on Matrix Field Theory},
    \href{https://doi.org/10.1007/978-3-319-46003-1}{\emph{Springer} (2016)}
    [\href{https://arxiv.org/abs/1603.00924}{\tt arXiv:1603.00924 [hep-th]}]
    [\href{https://inspirehep.net/search?p=find+EPRINT+arXiv:1603.00924}{\inspire}]

\bibitem{gauge1} 
    M. Buri\'{c}, L. Nenadovi\'{c}, D. Prekrat, 
    \emph{One-loop structure of the $U(1)$ gauge model on the truncated Heisenberg space},
    \href{https://doi.org/10.1140/epjc/s10052-016-4522-x}{\emph{EPJ} {\bf C 76} (2016) 672}
    [\href{https://arxiv.org/abs/1610.01429}{\tt arXiv:1610.01429 [hep-th]}]
    [\href{https://inspirehep.net/search?p=find+EPRINT+arXiv:1610.01429}{\inspire}]
    
\bibitem{gauge2} 
   L. Nenadovi\'{c}, 
    \emph{Properties of classical and quantum field theory on a curved noncommutative space}, 
    \href{http://nardus.mpn.gov.rs/handle/123456789/9453}{Ph.D. thesis (2017)} 
    
\bibitem{PP}
    E. Br\'{e}zin, C. Itzykson, G. Parisi, J.B. Zuber,
    \emph{Planar Diagrams}
    \href{https://doi.org/10.1007/BF01614153}{\emph{Commun. Math. Phys.} {\bf 59} (1978) 35}
    [\href{https://inspirehep.net/literature/122559}{\inspire}]
    
\bibitem{PPP}
    D. J. Gross, E. Witten,
    \emph{Possible Third Order Phase Transition in the Large $N$ Lattice Gauge Theory}
    \href{https://doi.org/10.1103/PhysRevD.21.446}{\emph{Phys. Rev.} {\bf D 21} (1980) 446-453}
    [\href{https://inspirehep.net/literature/157418}{\inspire}]


\bibitem{A00}
    V. G. Filev, D. O’Connor
    \emph{On the Phase Structure of Commuting Matrix Models},  
    \href{https://doi.org/10.1007/JHEP08(2014)003}{\emph{JHEP} {\bf 08} (2014) 003} 
    [\href{https://arxiv.org/abs/1402.2476v2}{\tt arXiv:1402.2476v2 [hep-th]}]
    [\href{https://inspirehep.net/literature/1280951}{\inspire}]


\bibitem{A01}
    D. O'Connor, C. S\"{a}mann, 
    \emph{Fuzzy scalar field theory as a multitrace matrix model},
    \href{https://doi.org/10.1088/1126-6708/2007/08/066}{\emph{JHEP} {\bf 08} (2007) 066} 
    [\href{https://arxiv.org/abs/0706.2493}{\tt arXiv:0706.2493 [hep-th]}]
    [\href{https://inspirehep.net/search?p=find+EPRINT+arXiv:0706.2493}{\inspire}]

\bibitem{A02}
    A. P. Polychronakos, 
    \emph{Effective action and phase transitions of scalar field on the fuzzy sphere}, 
    \href{https://doi.org/10.1103/PhysRevD.88.065010}{\emph{Phys. Rev.} {\bf D 88} (2013) 065010} 
    [\href{https://arxiv.org/abs/1306.6645}{\tt arXiv:1306.6645 [hep-th]}]
    [\href{https://inspirehep.net/search?p=find+EPRINT+arXiv:1306.6645}{\inspire}]

\bibitem{analytical2} 
    J. Tekel,  
    \emph{Uniform order phase and phase diagram of scalar field theory on fuzzy $\mathbb{C}P^n$},
    \href{https://doi.org/10.1007/JHEP10(2014)144}{\emph{JHEP} {\bf 10} (2014) 144}
    [\href{https://arxiv.org/abs/1407.4061}{\tt arXiv:1407.4061 [hep-th]}]
    [\href{https://inspirehep.net/search?p=find+EPRINT+arXiv:1407.4061}{\inspire}]

\bibitem{A03} 
    J. Tekel, 
    \emph{Matrix model approximations of fuzzy scalar field theories and their phase diagrams},
    \href{https://doi.org/10.1007/JHEP12(2015)176}{\emph{JHEP} {\bf 12} (2015) 176} 
    [\href{https://arxiv.org/abs/1510.07496}{\tt arXiv:1510.07496 [hep-th]}]
    [\href{https://inspirehep.net/search?p=find+EPRINT+arXiv:1510.07496}{\inspire}]

\bibitem{A04}
    J. Tekel, 
    \emph{Phase structure of fuzzy field theories and multitrace matrix models}, 
    \emph{Acta Phys. Slov.} {\bf 65} (2015) 369 
    [\href{https://arxiv.org/abs/1512.00689}{\tt arXiv:1512.00689 [hep-th]}]
    [\href{https://inspirehep.net/search?p=find+EPRINT+arXiv:1512.00689}{\inspire}]

\bibitem{analytical1} 
    J. Tekel,  
    \emph{Asymmetric hermitian matrix models and fuzzy field theory},
    \href{https://doi.org/10.1103/PhysRevD.97.125018}{\emph{Phys. Rev.} {\bf D 97} (2018) 125018}
    [\href{https://arxiv.org/abs/1711.02008}{\tt arXiv:1711.02008 [hep-th]}]
    [\href{https://inspirehep.net/search?p=find+EPRINT+arXiv:1711.02008}{\inspire}]  
    
\bibitem{A05}
    S. Rea and C. S\"{a}mann, 
    \emph{The Phase Diagram of Scalar Field Theory on the Fuzzy Disc},
    \href{https://doi.org/10.1007/JHEP11(2015)115}{\emph{JHEP} {\bf 11} (2015) 115}
    [\href{https://arxiv.org/abs/1507.05978}{\tt arXiv:1507.05978 [hep-th]}]
    [\href{https://inspirehep.net/search?p=find+EPRINT+arXiv:1507.05978}{\inspire}]

\bibitem{M2O}
    M. Xavier,
    \emph{A matrix phase for the $\phi^4$ scalar field on the fuzzy sphere},
    \href{https://doi.org/10.1088/1126-6708/2004/04/077}{\emph{JHEP} {\bf 04} (2004) 077}
    [\href{https://arxiv.org/abs/hep-th/0402230}{\tt arXiv:hep-th/0402230}]
    [\href{https://inspirehep.net/search?p=find+EPRINT+hep-th/0402230}{\inspire}]
    
\bibitem{N01}
    M. Panero, 
    \emph{Numerical simulations of a non-commutative theory: The Scalar model on the fuzzy sphere}, 
    \href{https://doi.org/10.1088/1126-6708/2007/05/082}{\emph{JHEP} {\bf 05} (2007) 082}
    [\href{https://arxiv.org/abs/hep-th/0608202}{\tt arXiv:hep-th/0608202}]
    [\href{https://inspirehep.net/search?p=find+EPRINT+hep-th/0608202}{\inspire}]
    
\bibitem{N02}
    F. Garcia Flores, X. Martin, D. O'Connor, 
    \emph{Simulation of a scalar field on a fuzzy sphere},
    \href{https://doi.org/10.1142/S0217751X09043195}{\emph{Int. J. Mod. Phys.} {\bf A 24} (2009) 3917}
    [\href{https://arxiv.org/abs/0903.1986}{\tt arXiv:0903.1986 [hep-lat]}]
    [\href{https://inspirehep.net/search?p=find+EPRINT+arXiv:0903.1986}{\inspire}]

\bibitem{N03}
    B. Ydri, 
    \emph{New algorithm and phase diagram of noncommutative $\phi^4$ on the fuzzy sphere}, 
    \href{https://doi.org/10.1007/JHEP03(2014)065}{\emph{JHEP} {\bf 03} (2014) 065}
    [\href{https://arxiv.org/abs/1401.1529}{\tt arXiv:1401.1529 [hep-th]}]
    [\href{https://inspirehep.net/search?p=find+EPRINT+arXiv:1401.1529}{\inspire}]

\bibitem{N04}
    B. Ydri, 
    \emph{Computational Physics: An Introduction to Monte Carlo Simulations of Matrix Field Theory}, 
    \href{https://doi.org/10.1142/10283}{\emph{World Scientific} (2017)}
    [\href{https://arxiv.org/abs/1506.02567}{\tt arXiv:1506.02567 [hep-lat]}]
    [\href{https://inspirehep.net/search?p=find+EPRINT+arXiv:1506.02567}{\inspire}]    
    
\bibitem{N05}
    B. Ydri, K. Ramda and A. Rouag, 
    \emph{Phase diagrams of the multitrace quartic matrix models of noncommutative $\Phi^4$ theory}, 
    \href{https://doi.org/10.1103/PhysRevD.93.065056}{\emph{Phys. Rev.} {\bf D 93} (2016) 065056}
    [\href{https://arxiv.org/abs/1509.03726}{\tt arXiv:1509.03726 [hep-th]}]
    [\href{https://inspirehep.net/search?p=find+EPRINT+arXiv:1509.03726}{\inspire}]

\bibitem{N06}
    P. Sabella-Garnier, 
    \emph{Time dependence of entanglement entropy on the fuzzy sphere}, 
    \href{https://doi.org/10.1007/JHEP08(2017)121}{\emph{JHEP} {\bf 08} (2017) 121}
    [\href{https://arxiv.org/abs/1705.01969}{\tt arXiv:1705.01969 [hep-th]}]
    [\href{https://inspirehep.net/search?p=find+EPRINT+arXiv:1705.01969}{\inspire}]
    
\bibitem{N07}
    M. P. Vachovski, 
    \emph{Numerical studies of the critical behaviour of non-commutative field theories},
    \href{http://mural.maynoothuniversity.ie/5439/}{Ph.D. thesis (2013)}

\bibitem{triple} 
    S. Kov\'{a}\v{c}ik, D. O'Connor,  
    \emph{Triple point of a scalar field theory on a fuzzy sphere},
    \href{https://doi.org/10.1007/JHEP10(2018)010}{\emph{JHEP} {\bf 10} (2018) 010}
    [\href{https://arxiv.org/abs/1805.08111}{\tt arXiv:1805.08111 [hep-th]}]
    [\href{https://inspirehep.net/search?p=find+EPRINT+arXiv:1805.08111}{\inspire}]

\bibitem{RG1}
    E. Br\'{e}zin, J. Zinn-Justin,
    \emph{Renormalization group approach to matrix models}
    \href{https://doi.org/10.1016/0370-2693(92)91953-7}{\emph{Phys. Lett.} {\bf B 288} (1992) 54}
    [\href{https://arxiv.org/abs/hep-th/9206035}{\tt arXiv:hep-th/9206035}]
    [\href{https://inspirehep.net/search?p=find+EPRINT+hep-th/9206035}{\inspire}]
     
\bibitem{RG2}
    B. Ydri, 
    \emph{The one-plaquette model limit of NC gauge theory in $2D$}
    \href{https://doi.org/10.1016/j.nuclphysb.2006.10.030}{\emph{Nucl. Phys.} {\bf B 762} (2007) 148}
    [\href{https://arxiv.org/abs/hep-th/0606206}{\tt arXiv:hep-th/0606206}]
    [\href{https://inspirehep.net/search?p=find+EPRINT+hep-th/0606206}{\inspire}]
     
\bibitem{RG3}
    S. Kawamoto, T. Kuroki, D. Tomino,
    \emph{Renormalization group approach to matrix models via noncommutative space}
    \href{https://doi.org/10.1007/JHEP08(2012)168}{\emph{JHEP} {\bf 08} (2012) 168}
    [\href{https://arxiv.org/abs/1206.0574}{\tt arXiv:1206.0574 [hep-th]}]
    [\href{https://inspirehep.net/search?p=find+EPRINT+arXiv:1206.0574}{\inspire}]

\bibitem{Ising}
    E. Ibarra-García-Padilla, C. G. Malanche-Flores, F. J. Poveda-Cuevas,
    \emph{The hobbyhorse of magnetic systems: the Ising model}
    \href{https://doi.org/10.1088/0143-0807/37/6/065103}{\emph{Eur. J. Phys.} {\bf 373} (2016) 6, 06510}
    [\href{https://arxiv.org/abs/1606.05800}{\tt arXiv:1606.05800 [cond-mat.stat-mech]}]
    [\href{https://inspirehep.net/literature/538490}{\inspire}]    

\bibitem{exponents}
    A. Pelissetto, E. Vicari,
    \emph{Critical Phenomena and Renormalization-GroupTheory}
    \href{https://doi.org/10.1016/S0370-1573(02)00219-3}{\emph{Phys. Rept.} {\bf 368} (2002) 549-727}
    [\href{https://arxiv.org/abs/cond-mat/0012164}{\tt arXiv:cond-mat/0012164 [cond-mat.stat-mech]}]
    [\href{https://inspirehep.net/literature/538490}{\inspire}]
    
\bibitem{hPT}
    W. Janke, D.A. Johnston, R. Kenna,
    \emph{Properties of higher-order phase transitions}
    \href{https://doi.org/10.1016/j.nuclphysb.2006.10.030}{\emph{Nucl. Phys.} {\bf B 736} (2006) 319-328}
    [\href{https://arxiv.org/abs/cond-mat/0512352}{\tt[arXiv:cond-mat/0512352 [cond-mat.stat-mech]]}]
    [\href{https://inspirehep.net/literature/711428}{\inspire}]

    
     	

\end{thebibliography}
\end{document}